%% file: main.tex
\documentclass{iopart}

\usepackage[nolist]{acronym}
\usepackage{amsopn}
\usepackage{cite}
\usepackage{color}
\usepackage{graphicx}
\usepackage{microtype}
\usepackage{xspace}
\usepackage{hyperref}
\expandafter\let\csname equation*\endcsname\relax
\expandafter\let\csname endequation*\endcsname\relax
\usepackage{amsmath}
\usepackage{amsfonts}
\usepackage{verbatim}
\usepackage{lineno}

\makeatletter
\newcommand{\cea}{Creswell et al\@ifnextchar.{}{.\@\xspace}}
\makeatother 

\newcommand*\patchAmsMathEnvironmentForLineno[1]{%
   \expandafter\let\csname old#1\expandafter\endcsname\csname #1\endcsname
   \expandafter\let\csname oldend#1\expandafter\endcsname\csname end#1\endcsname
   \renewenvironment{#1}%
      {\linenomath\csname old#1\endcsname}%
      {\csname oldend#1\endcsname\endlinenomath}}%
\newcommand*\patchBothAmsMathEnvironmentsForLineno[1]{%
   \patchAmsMathEnvironmentForLineno{#1}%
   \patchAmsMathEnvironmentForLineno{#1*}}%
\AtBeginDocument{%
\patchBothAmsMathEnvironmentsForLineno{equation}%
\patchBothAmsMathEnvironmentsForLineno{align}%
\patchBothAmsMathEnvironmentsForLineno{flalign}%
\patchBothAmsMathEnvironmentsForLineno{alignat}%
\patchBothAmsMathEnvironmentsForLineno{gather}%
\patchBothAmsMathEnvironmentsForLineno{multline}%
}

\begin{acronym}
\acro{GW}{gravitational wave}
\acro{LIGO}{Laser Interferometer Gravitational-wave Observatory}
\acro{LOSC}{\ac{LIGO} Open Science Center}
\acro{LSC}{\ac{LIGO} Scientific Collaboration}
\acro{LVC}{LIGO/Virgo Collaboration}
\acro{MAP}{maximum a posteriori}
\end{acronym}

\begin{document}

\title{A guide to LIGO-Virgo detector noise and extraction of transient gravitational-wave signals}

\input{authors_list}

\ead{\mailto{lvc.publications@ligo.org}}

\input{abstract}



\acresetall

\input{intro}

\input{data}

\input{noise}

\input{frequency}

\input{timefreq}

\input{further-reading}

\input{likelihood}
\input{search}

\input{pe}

\input{residuals}

\input{conclusion_new}

\ack
\input{LVCacknowledgments}


\section*{References}
\bibliographystyle{iopart-num}
\bibliography{refs}

\end{document}

%% file: authors_list.tex
\author{%
B~P~Abbott$^{1}$,  
R~Abbott$^{1}$,  
T~D~Abbott$^{2}$,  
S~Abraham$^{3}$,  
F~Acernese$^{4,5}$, 
K~Ackley$^{6}$,  
C~Adams$^{7}$,  
V~B~Adya$^{8,9}$,  
C~Affeldt$^{8,9}$,  
M~Agathos$^{10}$
% ...
and
J~Zweizig$^{1}$%
\\
{(The LIGO Scientific Collaboration and the Virgo Collaboration)}%
}%
\medskip
\address {${}^{\ast}$Deceased, February 2018.}%
\address {${}^{\dag}$Deceased, November 2017.}%
\address {${}^{\ddag}$Deceased, July 2018.}%
\medskip
\address {$^{1}$LIGO, California Institute of Technology, Pasadena, CA 91125, USA }
\address {$^{2}$Louisiana State University, Baton Rouge, LA 70803, USA }
\address {$^{3}$Inter-University Centre for Astronomy and Astrophysics, Pune 411007, India }
\address {$^{4}$Universit\`a di Salerno, Fisciano, I-84084 Salerno, Italy }
\address {$^{5}$INFN, Sezione di Napoli, Complesso Universitario di Monte S.Angelo, I-80126 Napoli, Italy }
\address {$^{6}$OzGrav, School of Physics \& Astronomy, Monash University, Clayton 3800, Victoria, Australia }
\address {$^{7}$LIGO Livingston Observatory, Livingston, LA 70754, USA }
\address {$^{8}$Max Planck Institute for Gravitational Physics (Albert Einstein Institute), D-30167 Hannover, Germany }
\address {$^{9}$Leibniz Universit\"at Hannover, D-30167 Hannover, Germany }
\address {$^{10}$University of Cambridge, Cambridge CB2 1TN, United Kingdom }
\address {$^{11}$University of Birmingham, Birmingham B15 2TT, United Kingdom }
\address {$^{12}$LIGO, Massachusetts Institute of Technology, Cambridge, MA 02139, USA }
\address {$^{13}$Instituto Nacional de Pesquisas Espaciais, 12227-010 S\~{a}o Jos\'{e} dos Campos, S\~{a}o Paulo, Brazil }
\address {$^{14}$Gran Sasso Science Institute (GSSI), I-67100 L'Aquila, Italy }
\address {$^{15}$INFN, Laboratori Nazionali del Gran Sasso, I-67100 Assergi, Italy }
\address {$^{16}$International Centre for Theoretical Sciences, Tata Institute of Fundamental Research, Bengaluru 560089, India }
\address {$^{17}$NCSA, University of Illinois at Urbana-Champaign, Urbana, IL 61801, USA }
\address {$^{18}$Universit\`a di Pisa, I-56127 Pisa, Italy }
\address {$^{19}$INFN, Sezione di Pisa, I-56127 Pisa, Italy }
\address {$^{20}$Departamento de Astronom\'{\i }a y Astrof\'{\i }sica, Universitat de Val\`encia, E-46100 Burjassot, Val\`encia, Spain }
\address {$^{21}$OzGrav, Australian National University, Canberra, Australian Capital Territory 0200, Australia }
\address {$^{22}$Laboratoire des Mat\'eriaux Avanc\'es (LMA), CNRS/IN2P3, F-69622 Villeurbanne, France }
\address {$^{23}$University of Wisconsin-Milwaukee, Milwaukee, WI 53201, USA }
\address {$^{24}$SUPA, University of Strathclyde, Glasgow G1 1XQ, United Kingdom }
\address {$^{25}$LAL, Univ. Paris-Sud, CNRS/IN2P3, Universit\'e Paris-Saclay, F-91898 Orsay, France }
\address {$^{26}$California State University Fullerton, Fullerton, CA 92831, USA }
\address {$^{27}$APC, AstroParticule et Cosmologie, Universit\'e Paris Diderot, CNRS/IN2P3, CEA/Irfu, Observatoire de Paris, Sorbonne Paris Cit\'e, F-75205 Paris Cedex 13, France }
\address {$^{28}$European Gravitational Observatory (EGO), I-56021 Cascina, Pisa, Italy }
\address {$^{29}$Chennai Mathematical Institute, Chennai 603103, India }
\address {$^{30}$Universit\`a di Roma Tor Vergata, I-00133 Roma, Italy }
\address {$^{31}$INFN, Sezione di Roma Tor Vergata, I-00133 Roma, Italy }
\address {$^{32}$INFN, Sezione di Roma, I-00185 Roma, Italy }
\address {$^{33}$Laboratoire d'Annecy de Physique des Particules (LAPP), Univ. Grenoble Alpes, Universit\'e Savoie Mont Blanc, CNRS/IN2P3, F-74941 Annecy, France }
\address {$^{34}$Embry-Riddle Aeronautical University, Prescott, AZ 86301, USA }
\address {$^{35}$Montclair State University, Montclair, NJ 07043, USA }
\address {$^{36}$Max Planck Institute for Gravitational Physics (Albert Einstein Institute), D-14476 Potsdam-Golm, Germany }
\address {$^{37}$Nikhef, Science Park 105, 1098 XG Amsterdam, The Netherlands }
\address {$^{38}$Korea Institute of Science and Technology Information, Daejeon 34141, South Korea }
\address {$^{39}$West Virginia University, Morgantown, WV 26506, USA }
\address {$^{40}$Universit\`a di Perugia, I-06123 Perugia, Italy }
\address {$^{41}$INFN, Sezione di Perugia, I-06123 Perugia, Italy }
\address {$^{42}$Syracuse University, Syracuse, NY 13244, USA }
\address {$^{43}$University of Minnesota, Minneapolis, MN 55455, USA }
\address {$^{44}$SUPA, University of Glasgow, Glasgow G12 8QQ, United Kingdom }
\address {$^{45}$LIGO Hanford Observatory, Richland, WA 99352, USA }
\address {$^{46}$Caltech CaRT, Pasadena, CA 91125, USA }
\address {$^{47}$Wigner RCP, RMKI, H-1121 Budapest, Konkoly Thege Mikl\'os \'ut 29-33, Hungary }
\address {$^{48}$University of Florida, Gainesville, FL 32611, USA }
\address {$^{49}$Stanford University, Stanford, CA 94305, USA }
\address {$^{50}$Universit\`a di Camerino, Dipartimento di Fisica, I-62032 Camerino, Italy }
\address {$^{51}$Universit\`a di Padova, Dipartimento di Fisica e Astronomia, I-35131 Padova, Italy }
\address {$^{52}$INFN, Sezione di Padova, I-35131 Padova, Italy }
\address {$^{53}$Montana State University, Bozeman, MT 59717, USA }
\address {$^{54}$Nicolaus Copernicus Astronomical Center, Polish Academy of Sciences, 00-716, Warsaw, Poland }
\address {$^{55}$OzGrav, University of Adelaide, Adelaide, South Australia 5005, Australia }
\address {$^{56}$Theoretisch-Physikalisches Institut, Friedrich-Schiller-Universit\"at Jena, D-07743 Jena, Germany }
\address {$^{57}$INFN, Sezione di Milano Bicocca, Gruppo Collegato di Parma, I-43124 Parma, Italy }
\address {$^{58}$Rochester Institute of Technology, Rochester, NY 14623, USA }
\address {$^{59}$Center for Interdisciplinary Exploration \& Research in Astrophysics (CIERA), Northwestern University, Evanston, IL 60208, USA }
\address {$^{60}$INFN, Sezione di Genova, I-16146 Genova, Italy }
\address {$^{61}$RRCAT, Indore, Madhya Pradesh 452013, India }
\address {$^{62}$Faculty of Physics, Lomonosov Moscow State University, Moscow 119991, Russia }
\address {$^{63}$OzGrav, University of Western Australia, Crawley, Western Australia 6009, Australia }
\address {$^{64}$Department of Astrophysics/IMAPP, Radboud University Nijmegen, P.O. Box 9010, 6500 GL Nijmegen, The Netherlands }
\address {$^{65}$Artemis, Universit\'e C\^ote d'Azur, Observatoire C\^ote d'Azur, CNRS, CS 34229, F-06304 Nice Cedex 4, France }
\address {$^{66}$Physik-Institut, University of Zurich, Winterthurerstrasse 190, 8057 Zurich, Switzerland }
\address {$^{67}$Univ Rennes, CNRS, Institut FOTON - UMR6082, F-3500 Rennes, France }
\address {$^{68}$Cardiff University, Cardiff CF24 3AA, United Kingdom }
\address {$^{69}$Washington State University, Pullman, WA 99164, USA }
\address {$^{70}$University of Oregon, Eugene, OR 97403, USA }
\address {$^{71}$Laboratoire Kastler Brossel, Sorbonne Universit\'e, CNRS, ENS-Universit\'e PSL, Coll\`ege de France, F-75005 Paris, France }
\address {$^{72}$Universit\`a degli Studi di Urbino 'Carlo Bo,' I-61029 Urbino, Italy }
\address {$^{73}$INFN, Sezione di Firenze, I-50019 Sesto Fiorentino, Firenze, Italy }
\address {$^{74}$Astronomical Observatory Warsaw University, 00-478 Warsaw, Poland }
\address {$^{75}$VU University Amsterdam, 1081 HV Amsterdam, The Netherlands }
\address {$^{76}$University of Maryland, College Park, MD 20742, USA }
\address {$^{77}$School of Physics, Georgia Institute of Technology, Atlanta, GA 30332, USA }
\address {$^{78}$Universit\'e Claude Bernard Lyon 1, F-69622 Villeurbanne, France }
\address {$^{79}$Universit\`a di Napoli 'Federico II,' Complesso Universitario di Monte S.Angelo, I-80126 Napoli, Italy }
\address {$^{80}$NASA Goddard Space Flight Center, Greenbelt, MD 20771, USA }
\address {$^{81}$Dipartimento di Fisica, Universit\`a degli Studi di Genova, I-16146 Genova, Italy }
\address {$^{82}$RESCEU, University of Tokyo, Tokyo, 113-0033, Japan. }
\address {$^{83}$Tsinghua University, Beijing 100084, China }
\address {$^{84}$Texas Tech University, Lubbock, TX 79409, USA }
\address {$^{85}$The University of Mississippi, University, MS 38677, USA }
\address {$^{86}$Museo Storico della Fisica e Centro Studi e Ricerche ``Enrico Fermi'', I-00184 Roma, Italyrico Fermi, I-00184 Roma, Italy }
\address {$^{87}$The Pennsylvania State University, University Park, PA 16802, USA }
\address {$^{88}$National Tsing Hua University, Hsinchu City, 30013 Taiwan, Republic of China }
\address {$^{89}$Charles Sturt University, Wagga Wagga, New South Wales 2678, Australia }
\address {$^{90}$Canadian Institute for Theoretical Astrophysics, University of Toronto, Toronto, Ontario M5S 3H8, Canada }
\address {$^{91}$University of Chicago, Chicago, IL 60637, USA }
\address {$^{92}$The Chinese University of Hong Kong, Shatin, NT, Hong Kong }
\address {$^{93}$Seoul National University, Seoul 08826, South Korea }
\address {$^{94}$Pusan National University, Busan 46241, South Korea }
\address {$^{95}$Carleton College, Northfield, MN 55057, USA }
\address {$^{96}$INAF, Osservatorio Astronomico di Padova, I-35122 Padova, Italy }
\address {$^{97}$INFN, Trento Institute for Fundamental Physics and Applications, I-38123 Povo, Trento, Italy }
\address {$^{98}$OzGrav, University of Melbourne, Parkville, Victoria 3010, Australia }
\address {$^{99}$Columbia University, New York, NY 10027, USA }
\address {$^{100}$Universitat de les Illes Balears, IAC3---IEEC, E-07122 Palma de Mallorca, Spain }
\address {$^{101}$Universit\'e Libre de Bruxelles, Brussels 1050, Belgium }
\address {$^{102}$Sonoma State University, Rohnert Park, CA 94928, USA }
\address {$^{103}$Departamento de Matem\'aticas, Universitat de Val\`encia, E-46100 Burjassot, Val\`encia, Spain }
\address {$^{104}$University of Rhode Island, Kingston, RI 02881, USA }
\address {$^{105}$The University of Texas Rio Grande Valley, Brownsville, TX 78520, USA }
\address {$^{106}$Bellevue College, Bellevue, WA 98007, USA }
\address {$^{107}$MTA-ELTE Astrophysics Research Group, Institute of Physics, E\"otv\"os University, Budapest 1117, Hungary }
\address {$^{108}$Institute for Plasma Research, Bhat, Gandhinagar 382428, India }
\address {$^{109}$The University of Sheffield, Sheffield S10 2TN, United Kingdom }
\address {$^{110}$IGFAE, Campus Sur, Universidade de Santiago de Compostela, 15782 Spain }
\address {$^{111}$Dipartimento di Scienze Matematiche, Fisiche e Informatiche, Universit\`a di Parma, I-43124 Parma, Italy }
\address {$^{112}$California State University, Los Angeles, 5151 State University Dr, Los Angeles, CA 90032, USA }
\address {$^{113}$Universit\`a di Trento, Dipartimento di Fisica, I-38123 Povo, Trento, Italy }
\address {$^{114}$Universit\`a di Roma 'La Sapienza,' I-00185 Roma, Italy }
\address {$^{115}$Colorado State University, Fort Collins, CO 80523, USA }
\address {$^{116}$Kenyon College, Gambier, OH 43022, USA }
\address {$^{117}$Christopher Newport University, Newport News, VA 23606, USA }
\address {$^{118}$National Astronomical Observatory of Japan, 2-21-1 Osawa, Mitaka, Tokyo 181-8588, Japan }
\address {$^{119}$Observatori Astron\`omic, Universitat de Val\`encia, E-46980 Paterna, Val\`encia, Spain }
\address {$^{120}$School of Mathematics, University of Edinburgh, Edinburgh EH9 3FD, United Kingdom }
\address {$^{121}$Institute Of Advanced Research, Gandhinagar 382426, India }
\address {$^{122}$Indian Institute of Technology Bombay, Powai, Mumbai 400 076, India }
\address {$^{123}$University of Szeged, D\'om t\'er 9, Szeged 6720, Hungary }
\address {$^{124}$Tata Institute of Fundamental Research, Mumbai 400005, India }
\address {$^{125}$INAF, Osservatorio Astronomico di Capodimonte, I-80131, Napoli, Italy }
\address {$^{126}$University of Michigan, Ann Arbor, MI 48109, USA }
\address {$^{127}$American University, Washington, D.C. 20016, USA }
\address {$^{128}$GRAPPA, Anton Pannekoek Institute for Astronomy and Institute of High-Energy Physics, University of Amsterdam, Science Park 904, 1098 XH Amsterdam, The Netherlands }
\address {$^{129}$Delta Institute for Theoretical Physics, Science Park 904, 1090 GL Amsterdam, The Netherlands }
\address {$^{130}$Directorate of Construction, Services \& Estate Management, Mumbai 400094 India }
\address {$^{131}$University of Bia{\l }ystok, 15-424 Bia{\l }ystok, Poland }
\address {$^{132}$King's College London, University of London, London WC2R 2LS, United Kingdom }
\address {$^{133}$University of Southampton, Southampton SO17 1BJ, United Kingdom }
\address {$^{134}$University of Washington Bothell, Bothell, WA 98011, USA }
\address {$^{135}$Institute of Applied Physics, Nizhny Novgorod, 603950, Russia }
\address {$^{136}$Ewha Womans University, Seoul 03760, South Korea }
\address {$^{137}$Inje University Gimhae, South Gyeongsang 50834, South Korea }
\address {$^{138}$National Institute for Mathematical Sciences, Daejeon 34047, South Korea }
\address {$^{139}$Ulsan National Institute of Science and Technology, Ulsan 44919, South Korea }
\address {$^{140}$Universit\"at Hamburg, D-22761 Hamburg, Germany }
\address {$^{141}$Maastricht University, P.O. Box 616, 6200 MD Maastricht, The Netherlands }
\address {$^{142}$NCBJ, 05-400 \'Swierk-Otwock, Poland }
\address {$^{143}$Institute of Mathematics, Polish Academy of Sciences, 00656 Warsaw, Poland }
\address {$^{144}$Cornell University, Ithaca, NY 14850, USA }
\address {$^{145}$Hillsdale College, Hillsdale, MI 49242, USA }
\address {$^{146}$Hanyang University, Seoul 04763, South Korea }
\address {$^{147}$Korea Astronomy and Space Science Institute, Daejeon 34055, South Korea }
\address {$^{148}$NASA Marshall Space Flight Center, Huntsville, AL 35811, USA }
\address {$^{149}$Dipartimento di Matematica e Fisica, Universit\`a degli Studi Roma Tre, I-00146 Roma, Italy }
\address {$^{150}$INFN, Sezione di Roma Tre, I-00146 Roma, Italy }
\address {$^{151}$ESPCI, CNRS, F-75005 Paris, France }
\address {$^{152}$OzGrav, Swinburne University of Technology, Hawthorn VIC 3122, Australia }
\address {$^{153}$University of Portsmouth, Portsmouth, PO1 3FX, United Kingdom }
\address {$^{154}$Southern University and A\&M College, Baton Rouge, LA 70813, USA }
\address {$^{155}$College of William and Mary, Williamsburg, VA 23187, USA }
\address {$^{156}$Centre Scientifique de Monaco, 8 quai Antoine Ier, MC-98000, Monaco }
\address {$^{157}$Indian Institute of Technology Madras, Chennai 600036, India }
\address {$^{158}$INFN Sezione di Torino, Via P.~Giuria 1, I-10125 Torino, Italy }
\address {$^{159}$Institut des Hautes Etudes Scientifiques, F-91440 Bures-sur-Yvette, France }
\address {$^{160}$IISER-Kolkata, Mohanpur, West Bengal 741252, India }
\address {$^{161}$Whitman College, 345 Boyer Avenue, Walla Walla, WA 99362 USA }
\address {$^{162}$Universit\'e de Lyon, F-69361 Lyon, France }
\address {$^{163}$Hobart and William Smith Colleges, Geneva, NY 14456, USA }
\address {$^{164}$Janusz Gil Institute of Astronomy, University of Zielona G\'ora, 65-265 Zielona G\'ora, Poland }
\address {$^{165}$University of Washington, Seattle, WA 98195, USA }
\address {$^{166}$SUPA, University of the West of Scotland, Paisley PA1 2BE, United Kingdom }
\address {$^{167}$Indian Institute of Technology, Gandhinagar Ahmedabad Gujarat 382424, India }
\address {$^{168}$Universit\'e de Montr\'eal/Polytechnique, Montreal, Quebec H3T 1J4, Canada }
\address {$^{169}$Indian Institute of Technology Hyderabad, Sangareddy, Khandi, Telangana 502285, India }
\address {$^{170}$International Institute of Physics, Universidade Federal do Rio Grande do Norte, Natal RN 59078-970, Brazil }
\address {$^{171}$Villanova University, 800 Lancaster Ave, Villanova, PA 19085, USA }
\address {$^{172}$Andrews University, Berrien Springs, MI 49104, USA }
\address {$^{173}$Max Planck Institute for Gravitationalphysik (Albert Einstein Institute), D-14476 Potsdam-Golm, Germany }
\address {$^{174}$Universit\`a di Siena, I-53100 Siena, Italy }
\address {$^{175}$Trinity University, San Antonio, TX 78212, USA }
\address {$^{176}$Van Swinderen Institute for Particle Physics and Gravity, University of Groningen, Nijenborgh 4, 9747 AG Groningen, The Netherlands }

%% file: abstract.tex
\begin{abstract}

The LIGO Scientific Collaboration and the Virgo Collaboration have cataloged eleven confidently detected gravitational-wave events during the first two observing runs of the advanced detector era.
All eleven events were consistent with being from well-modeled mergers between compact stellar-mass objects: black holes or neutron stars.
The data around the time of each of these events have been made publicly available through the
Gravitational-Wave Open Science Center.
The entirety of the gravitational-wave strain data from the first and second observing runs have also now been made publicly available.
There is considerable interest among the broad scientific community in understanding the 
data and methods used in the analyses. In this paper, we provide an overview of the detector noise properties 
and the data analysis techniques used to detect gravitational-wave signals and infer the source properties. 
We describe some of the checks that are performed to validate the analyses and results from the observations of gravitational-wave events.
We also address concerns that have been raised about various properties of LIGO-Virgo detector noise and the correctness of our analyses as applied to the resulting data. 
\end{abstract}

%% file: intro.tex
\section{Introduction}

Gravitational-wave observations have become an important new means to learn about the Universe. The LIGO Scientific Collaboration and the Virgo Collaboration (LVC) have published a series of discoveries
beginning with the first detected event, GW150914 \cite{Abbott:2016blz}, a binary black hole merger. 
Within a span of two years, that event was followed by nine other binary black hole detections (GW151012~\cite{TheLIGOScientific:2016qqj,LIGOScientific:2018mvr}, GW151226~\cite{Abbott:2016nmj}, GW170104~\cite{Abbott:2017vtc}, GW170608~\cite{Abbott:2017gyy}, GW170729, GW170809, GW170814~\cite{Abbott:2017oio}, GW170818 and GW170823),
and one binary neutron star merger, GW170817~\cite{TheLIGOScientific:2017qsa}.
Details about all of these confidently-detected gravitational-wave events have been published in a catalog, GWTC-1~\cite{LIGOScientific:2018mvr}.

The global gravitational-wave detector network currently consists of two Advanced LIGO detectors in the U.S.~\cite{TheLIGOScientific:2014jea} in Hanford, Washington and Livingston, Louisiana; the Advanced Virgo detector in Cascina, Italy~\cite{TheVirgo:2014hva}; and the GEO 600 detector in Germany \cite{Dooley_2016}. In the coming years this network will grow through the addition of the Japanese detector, KAGRA~\cite{Aso:2013,Akutsu:2017thy,Akutsu:2019rba}, and a third Advanced LIGO detector to be located in India~\cite{Unnikrishnan:2013qwa}.
The first observing run (O1) of Advanced LIGO took place from September 12, 2015 until January 19, 2016. The second observing run (O2) for the Advanced  LIGO  detectors began on November 30, 2016, and lasted until August 25, 2017.  The Advanced Virgo detector formally commenced observations in O2 on August 1, 2017, enabling the first three-detector observations of gravitational waves~\cite{LIGOScientific:2018mvr}. A third LIGO-Virgo observing run, O3, began on April 1, 2019, with all three detectors operating with their best sensitivity to date.

Consistency between multiple detectors helps greatly to suppress instrumental backgrounds and to allow coherent analysis of gravitational-wave signals.
All of the event detections published to date have involved both of the Advanced LIGO detectors, while
GW170814 and GW170818 were triple-detections sensed by Virgo as well.
Data from Virgo were also used in the parameter estimation analysis and sky localization determination for GW170729, GW170809, and GW170817.
The Virgo data were especially critical in helping to find the source of GW170817~\cite{2017ApJ...848L..12A}.
This binary neutron star merger represented a remarkable debut for multi-messenger astronomy with gravitational waves, as it was closely followed by a short gamma-ray burst, GRB 170817A~\cite{2017ApJ...848L..14G,Monitor:2017mdv}, and the relatively precise localization obtained from the gravitational-wave data enabled the identification and thorough multi-wavelength study of kilonova and afterglow emission from an optical counterpart, SSS17a / AT 2017gfo~\cite{2017Sci...358.1556C,2017ApJ...848L..12A}.

As summarized in~\cite{LIGOScientific:2018mvr}, the LVC detections were made using two independent matched-filter analyses to search for compact binary coalescences in O2~\cite{Usman:2015kfa,Messick:2016aqy}, as well as an unmodeled search for short-duration transient signals or \emph{bursts}~\cite{Klimenko:2015ypf}. 
Thus, detection methods that were developed by the LVC and tested using simulated signals added to mock data, or to previous sets of real data where any possible signals were overwhelmed by noise, have now been demonstrated to be effective for astrophysical gravitational-wave signals.
Testing and validation of LVC analyses was achieved using both (simulated) signal injections performed within the analysis, i.e.\ in software, and signal injections made in hardware by moving the detectors' test masses.

The growth of the number of observed gravitational-wave events has stimulated intense interest in the astrophysical implications of the detected sources, as well as interest in the gravitational-wave data. Currently, the LVC releases data through the Gravitational-Wave Open Science Center (GWOSC)~\cite{Vallisneri:2014vxa,LOSC}. 
LIGO data releases are described in the LIGO Data Management Plan~\cite{plan}, an agreement between the LIGO Laboratory and the US National Science Foundation. The LVC policy for releasing gravitational-wave triggers and event candidates is presented in ~\cite{plan-LVC,LV_alerts}.
For detections of compact binary mergers, about one hour (4096 seconds) of calibrated strain data around the event time are released at the time of publication.  These data are available for all published detections in O1 and O2~\cite{detections}.  

Currently the bulk data from the initial LIGO Science Runs since 2005 are available on GWOSC~\cite{LOSC}, as are the Advanced LIGO data from the O1 observing run~\cite{O1_GWOSC} and the Advanced LIGO and Advanced Virgo data from the O2 observing run~\cite{O2_GWOSC}.  Timing for release of data in future observing runs is described in the Data Management Plan~\cite{plan}; for instance, the bulk data from the first 6 months of the O3 run will be released in April 2021.  GWOSC is continually updating and releasing data products that address the needs and interests of the broader scientific community. Many of the analysis software packages used by the collaboration are publicly available as open source code; a list of these is available on the GWOSC web site~\cite{LOSC}.  Also, a number of intermediate data products are released through the LIGO Document Control Center, typically linked with LVC papers; e.g.\ see~\cite{LIGO-P1800370}. 

\begin{figure}[h]
\centering \includegraphics[width=\textwidth]{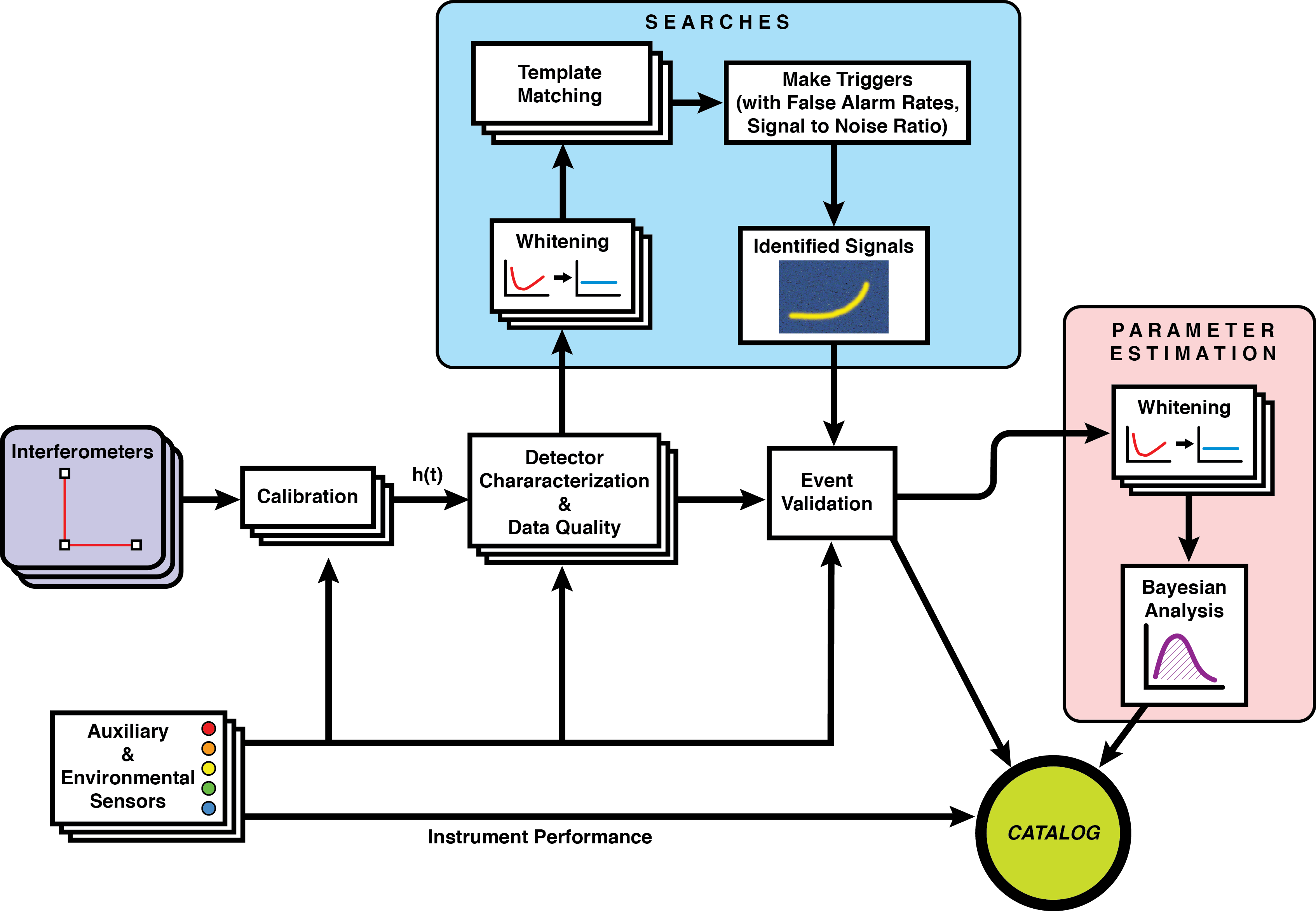} 
\caption{\label{fig:summ_diagram} A simplified schematic summarizing the main steps in LIGO-Virgo data processing, from the output of the data to the results reported in a catalog of transient events.}
\end{figure}

With the public release of the LIGO and Virgo data, groups outside these collaborations are analyzing the released data. 
Most of these analyses are producing results consistent with the LVC's~\cite{Green:2017voq,Nielsen:2018bhc,Roulet:2018jbe,Dai:2018dca,Radice:2018ozg,De:2018uhw,De:2018zrk,Nitz:2018imz}, and some additional significant event candidates have been reported~\cite{Zackay:2019-O1event,Venumadhav:2019-O2events}. 
The noise properties of the LIGO data and the correctness of the LVC data analysis for GW150914 have also been questioned~\cite{Creswell:2017rbh,Creswell:2018tsr}, although successive gravitational wave detections have strengthened confidence in our detection and parameter estimation methods~\cite{LIGOScientific:2018mvr}.
Motivated by the widespread interest in analyzing LIGO and Virgo data, in this paper we provide an overview of the properties of the LIGO-Virgo data and its noise components. We also describe the essential features of data analysis procedures that have been used by LIGO and Virgo teams to detect and measure the properties of the cataloged gravitational-wave sources~\cite{LIGOScientific:2018mvr}, as summarized in Figure~\ref{fig:summ_diagram}.
The analysis of LIGO and Virgo data in searching for gravitational-wave signals is complex, as is the correct treatment of the statistical properties of noise. The LVC encourages the broader scientific community to access and analyze its data, and will always be open to discussions about the methods it uses to arrive at its conclusions. 
The codes used to analyze LIGO-Virgo data are public. The special purpose codes used to generate many of the figures in this paper are also available~\cite{https://doi.org/10.5281/zenodo.3380539}.
In addition, the LVC has made available a Jupyter notebook to illustrate methods used to produce key figures and results in a simplified implementation~\cite{Tutorial_GWOSC}.
Finally, many of the software packages used by the LVC to process the LIGO-Virgo data, search for events and characterize observed signals can be found at the GWOSC site~\cite{Software_GWOSC}.

The paper is organized as follows. In Section~\ref{sec:data} we  describe the properties of the LIGO-Virgo data, while in Section~\ref{sec:noise} we discuss the noise that affects those data. Section~\ref{sec:freq} describes the basic data processing steps used to properly Fourier transform the data and estimate the power spectrum.  
Section~\ref{sec:timefreq} describes wavelet based time-frequency methods that can be used to assess possible deviations from stationary detector noise.  Section~\ref{s:further-reading} addresses detector and calibration issues for LIGO and Virgo.  Section~\ref{sec:likelihood} describes the noise model used to define the likelihood function used in parameter estimation studies. Section~\ref{s:searches} gives a description of the means by which the LVC searches for gravitational-wave signals, while Section~\ref{s:pe} presents the means by which the LVC infers the detected waveforms and estimates the physical parameters for the system that emitted the gravitational waves. 
To illustrate these concepts, Section~\ref{sec:res} provides a simplified description of how the publicly released data surrounding GW150914 can be used to find a best fit waveform model and to study the correlation properties of the residuals.
We also address claims made in~\cite{Creswell:2018tsr,Creswell:2017rbh} concerning correlations in detector noise, residuals, and the estimation of GW150914's source properties.
In addressing these claims, the LVC notes that it is beneficial for gravitational-wave science that groups external to our collaboration can introduce new ideas and techniques.
Finally, in Section~\ref{sec:conclusions} we provide a summary assessment of LIGO and Virgo data properties as well as LVC data analysis findings and validation.

%% file: data.tex
\begin{input}
\section{Properties of LIGO-Virgo data}
\label{sec:data}

The Advanced LIGO~\cite{TheLIGOScientific:2014jea} and Advanced Virgo~\cite{TheVirgo:2014hva} second-generation gravitational-wave detectors are large-scale enhanced Michelson interferometers.
The detectors are sensitive to space time strain induced by passing gravitational waves, as well as equivalent terrestrial force and displacement noises, each causing the lengths of the arms to vary over time. 
Differences in relative arm length generate power variations in the enhanced Michelson’s output, captured by photodiodes. 
The signal from these photodiodes serve as both the gravitational-wave readout and an error signal for controlling the relative arm length below roughly 100~Hz. 

The Advanced LIGO gravitational-wave detectors are identical in design, 
with 4~km long
arms. Advanced Virgo has a similar design, with 3~km long arms. 
Fabry-Perot cavities are used in the arms of the detectors to increase
the interaction time with a gravitational wave, and power recycling is used to
increase the effective laser power. 
Signal recycling has been added in the 
Advanced LIGO detectors to shape their frequency response~\cite{TheLIGOScientific:2014jea}. Advanced Virgo has not yet implemented signal recycling, but will in the future~\cite{TheVirgo:2014hva}. 

A calibration procedure is applied to the interferometer photodiode output of each detector (see Section \ref{ss:calibration}) to produce gravitational-wave strain data as a time series, sampled at 16384 Hz for LIGO data 
and 20~kHz for Virgo data. 
For the Advanced LIGO detectors, the calibration is valid above 10~Hz and below 5~kHz, as described in Section \ref{ss:calibration}. 
For Advanced Virgo in O2 the calibration validity range was from 10 Hz to 8 kHz~\cite{Acernese:2018bfl}.
The detectors also record hundreds of thousands of \textit{auxiliary channels}, time series recorded in addition to the strain signal, that monitor the behavior of the detectors and their environment. 
The GWOSC provides distilled additional channels of data in which flags pertaining to different levels of problems with the data quality are implemented~\footnote{https://www.gw-openscience.org/segments}. 
We employ continuous monitoring of the detector performance 
to characterize noise sources that could negatively impact the sensitivity of the searches or the source property estimation~\cite{TheLIGOScientific:2016zmo,Covas:2018}. Invalid data due to detector malfunction, calibration error, or data acquisition problems are flagged so that they can be removed from analyses, as described in Section \ref{s:further-reading} and~\cite{TheLIGOScientific:2017lwt}.

\end{input}

{\color{blue} [sll]: After the introduction, but before the section on Properties of Noise, I think we need another brief section that describes the data itself and outlines the analysis problem at a high level.  It can be/should be only a page or so, but I think is needed to provide the uninitiated some framework to understand what is being talked about in the rest of the paper. It should include:
\begin{itemize}
   \item Description of the form of the data -- how many channels, how many detectors how often is it sampled
   \item A cursory note about the fact that there is considerable environmental monitoring to characterize noise and other factors that might influence detector health and data quality
   \item A rough outline describing the scheme for analysis -- quick immediate significance analysis at each detector, followed by comparison between detectors, human interactions at what points, then how it is passed off to analysis
   \item A rough enumeration of what we are doing all the time to characterize what is going on in the detector -- simple statements like we are characterizing the noise (point to current section 2), detector charcterization, etc.
   \item A rough enumeration of the different analysis pipelines that are being run on the data, and when (e.g. we want to talk about low latency pipelines, also be explicit about multiple independent pipelines that all return the same events, and then also mention that we also have deep digging pipelines like what we just did to produce the catalog paper)
   \end{itemize}
As I said -- I don't think it needs to be exhaustive -- it just needs to be some framework that shows this is complex and has many moving parts, and we've done due diligence to address all those moving parts with plenty of internal crosschecks and oversight.}

%% file: noise.tex
\section{Basic properties of detector noise}
\label{sec:noise}
\bigskip

The data recorded by the Advanced LIGO and Advanced Virgo instruments are impacted by many sources of noise, including quantum sensing noise, seismic noise, suspension thermal noise, mirror coating thermal noise, and gravity gradient noise~\cite{TheLIGOScientific:2014jea}. 
In addition, there are transient noise events, for example coming from anthropogenic sources, weather, equipment malfunctions~\cite{TheLIGOScientific:2016zmo}, as well as occasional transient noise of unknown origin~\cite{Cabero:2019orq}.
There is also persistent elevated noise confined to certain frequencies, manifesting as very narrow peaks in a plot of noise versus frequency, which we refer to as spectral \emph{lines}; these are typically caused by electrical and mechanical devices or resonances~\cite{Covas:2018}.
The combination of all the noise sources in a detector produces a time series $n(t)$ that can be represented by a vector ${\bf n}$, with components given by the discrete time samples $n_i = n(t_i)$. The noise is described as a stochastic process with statistical properties given by the joint probability distribution $p({\bf n})$. This model can be used to define summary statistics such as the mean $\mu= {\rm E}[{\bf n}]$ (where ${\rm E}$ is defined as the expectation value) and covariance $C_{ij} = {\rm E}[(n_i - \mu)(n_j- \mu)]$
where the expectation values are taken with respect to $p({\bf n})$. The mean can be estimated from the data as
\begin{equation}
\hat{\mu} = \frac{1}{N}\sum_{i=1}^N n_i \, ,
\end{equation}
where $N={\rm dim}({\bf n})$ is the number of data samples. The full covariance matrix cannot be estimated from the data without making additional assumptions as we have only $M=1$ measurements for each data point, rendering the sample covariance matrix formally undefined:
\begin{equation}
\hat{C}_{ij} = \frac{1}{M-1} (n_i-\hat{\mu}) ( n_j-\hat{\mu}) \, .
\end{equation}
Estimates of the covariance matrix can be made if noise is assumed to follow a particular distribution, or if the noise properties are unchanging in time.
Note that in practice, analyses generally do not use all $N$ samples at once, but rather use segments of contiguous data of various lengths from a few seconds up to hours depending on the intended application. Noise is referred to as {\em Gaussian} if the joint probability distribution follows a multi-variate normal distribution:
\begin{equation}\label{gauss}
p({\bf n}) =  \frac{1}{ {\rm det}(2 \pi {\bf C})^{1/2}} \, \exp\left[- \frac{1}{2} \sum_{ij} (n_i - \mu)(n_j- \mu) C_{ij}^{-1}\right] \, ,
\end{equation}
where $C_{ij}^{-1}$ is the inverse of the covariance matrix at $i,j$.
The noise is referred to as {\em stationary} if $C_{ij}$ depends only on the lag $|i-j|$.  Stationary noise is characterized by the correlation function $C(\tau)$, where $\tau=|t_i-t_j|$ is the time lag.  
Transforming to the Fourier domain, where the labels $i, j$ now refer to frequencies $f_i, f_j$, stationary noise has a diagonal covariance matrix $C_{ij} = \delta_{ij} S_n(f_i)$, which defines the power spectral density $S_n(f)$. The power spectral density is given by the Fourier transform of the correlation function $C(\tau)$.
Amplitude spectral density is the square root of power spectral density and has units of Hz$^{-1/2}$.
The noise is referred to as {\em white} if $C_{ij} = \delta_{ij} \sigma^2$ in both the frequency domain and the time domain. 
White noise is, however, a poor approximation to LIGO-Virgo detector noise

Understanding the noise is crucial to detecting gravitational-wave signals and inferring the properties of the astrophysical sources that generate them. Improper modeling of the noise can result in the significance of an event being incorrectly estimated, and to systematic biases in the parameter estimation. To guard against these unwanted outcomes, detector characterization and noise modeling are significant activities within the LVC~\cite{TheLIGOScientific:2016zmo,TheLIGOScientific:2017lwt}. While many textbook treatments of gravitational-wave data analysis~\cite{alma991002489675403836, creighton2012gravitational, Jaranowski2012} describe the idealized case of independent detectors with stationary, Gaussian noise, actual LVC analyses are careful to account for deviations from this ideal. 

\begin{figure}[t]
\includegraphics[width=\textwidth]{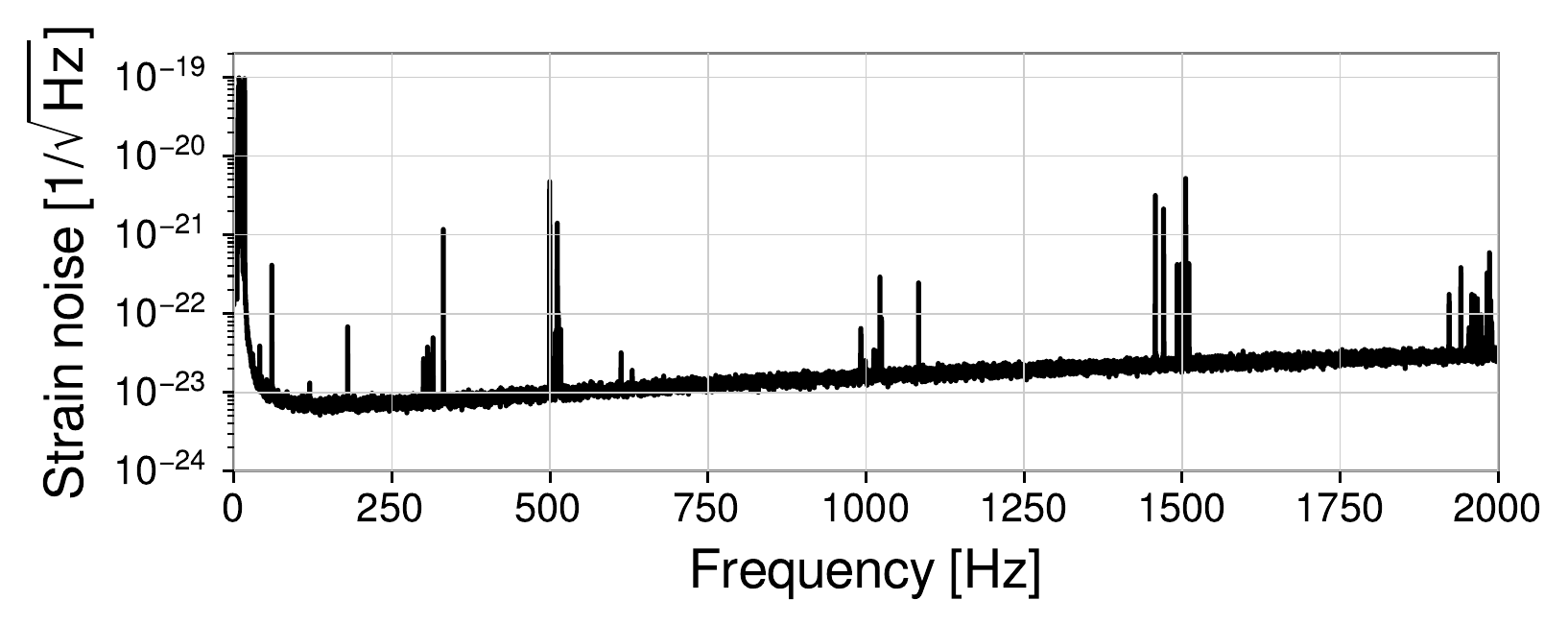}
\includegraphics[width=\textwidth]{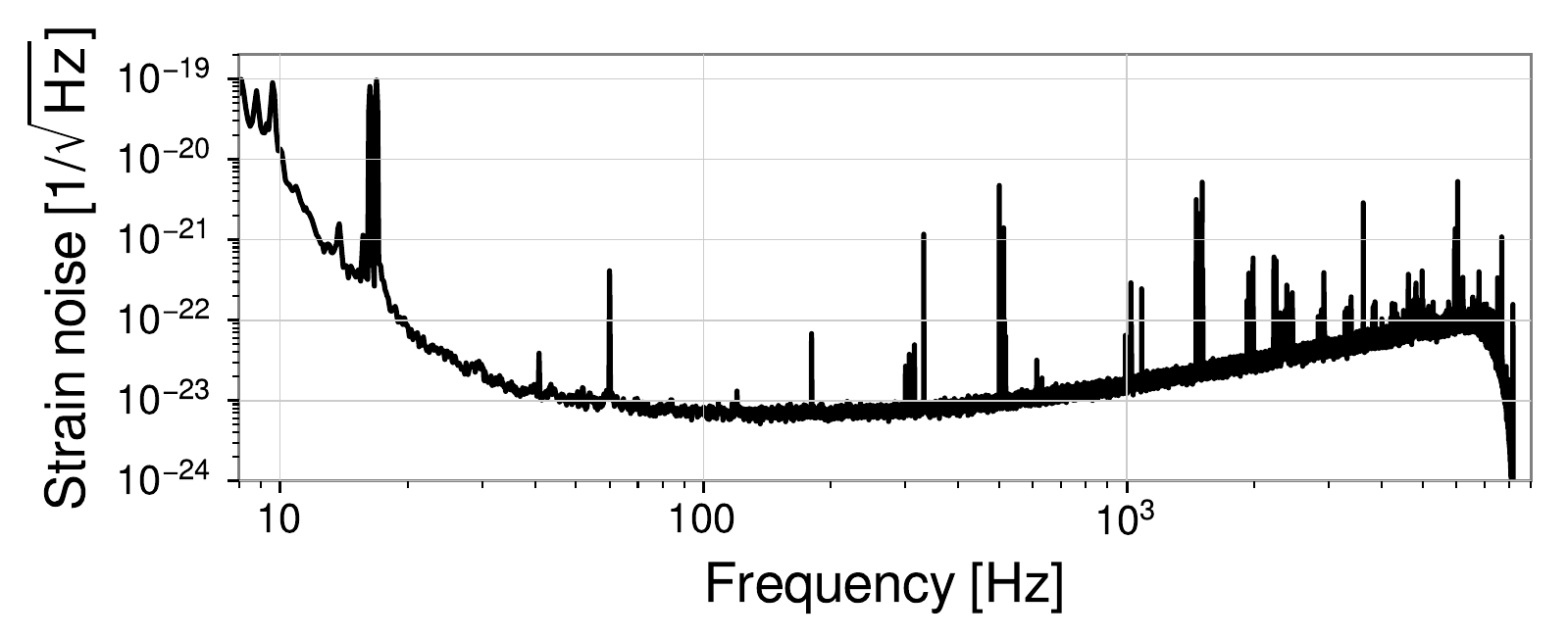}
\caption{Amplitude spectral density of the LIGO-Livingston detector data, using 10-second fast Fourier transforms and averaged over a three-minute period starting at August 17, 2017 12:36:00 UTC, five minutes before the merger time of GW170817~\cite{TheLIGOScientific:2017qsa}. The top plot is a linear frequency scale, highlighting periodic features from 0~Hz to 2000~Hz. The bottom plot is a log scale, illustrating the features in the detector data from 8~Hz to the Nyquist frequency 8192 Hz.
} 
\label{f:asds}
\end{figure}

The Advanced LIGO and Advanced Virgo detector data have a rich structure in both time and frequency.
For a given gravitational-wave source, the noise (as described by its spectral density) governs the measured signal-to-noise ratio (SNR). 
Figure \ref{f:asds} shows the spectral frequency content of the LIGO-Livingston detector averaged over a three-minute period shortly before the first detection of gravitational waves from a binary neutron star merger, GW170817.
During the O1 and O2 runs, the Advanced LIGO detectors had an averaged measured noise amplitude of about $10^{-23}$ Hz$^{-1/2}$ at 100 Hz.
(The target sensitivity at 100 Hz for Advanced LIGO is $4 \times 10^{-24}$ Hz$^{-1/2}$~\cite{TheLIGOScientific:2014jea}, while for Advanced Virgo it is $5 \times 10^{-24}$ Hz$^{-1/2}$~\cite{TheVirgo:2014hva}.)
The steep shape at low frequencies is dominated by noise related to ground motion. Above roughly 100~Hz, the Advanced LIGO detectors are currently quantum noise limited, and their noise curves are dominated by shot noise~\cite{TheLIGOScientific:2014jea, TheLIGOScientific:2016agk}. High amplitude noise features are also present in the data at certain frequencies, including lines due to the AC power grid (harmonics of 60~Hz in the U.S.\ and 50~Hz in Europe), mechanical resonances of the mirror suspensions, injected calibration lines, and noise entering through the detector control systems.
For a detailed account of noise sources that appear at specific frequencies in the Advanced LIGO detectors, see ~\cite{Covas:2018}. 
For a list of the Advanced Virgo noise lines for observing run O2, see~\cite{V1-lines-O2}.

%% file: frequency.tex
\section{Fourier domain analysis}
\label{sec:freq}

The noise in the LIGO-Virgo detectors is, with isolated exceptions, approximately stationary, and therefore can be most easily characterized in the frequency domain. Stationary, Gaussian noise is uncorrelated between frequency bins, and the noise $\tilde{n}(f)$ in each bin follows a Gaussian distribution with random phase and amplitude $S_n^{1/2}(f)$. The first step in many LVC analyses is to Fourier transform the time-domain data using a fast Fourier transform (FFT)~\cite{CooleyTukey,5217220,Rao:2010:FFT:1941838}.  Since the FFT implicitly assumes that the stretch of data being transformed is periodic in time, window functions~\cite{Tukey1967,1455106} have to be applied to the data to suppress spectral leakage~\cite{1455106} using {\it e.g.}\ a Tukey (cosine-tapered) window function. Failing to window the data will lead to spectral leakage and spurious correlations in the phase between bins.
For the analysis of transient data the use of Tukey windows is advantageous as signals will suffer less modification than, for example, Hanning or Flattop windows~\cite{1455106}.

\begin{figure}[h]
\centering \includegraphics[scale=0.23]{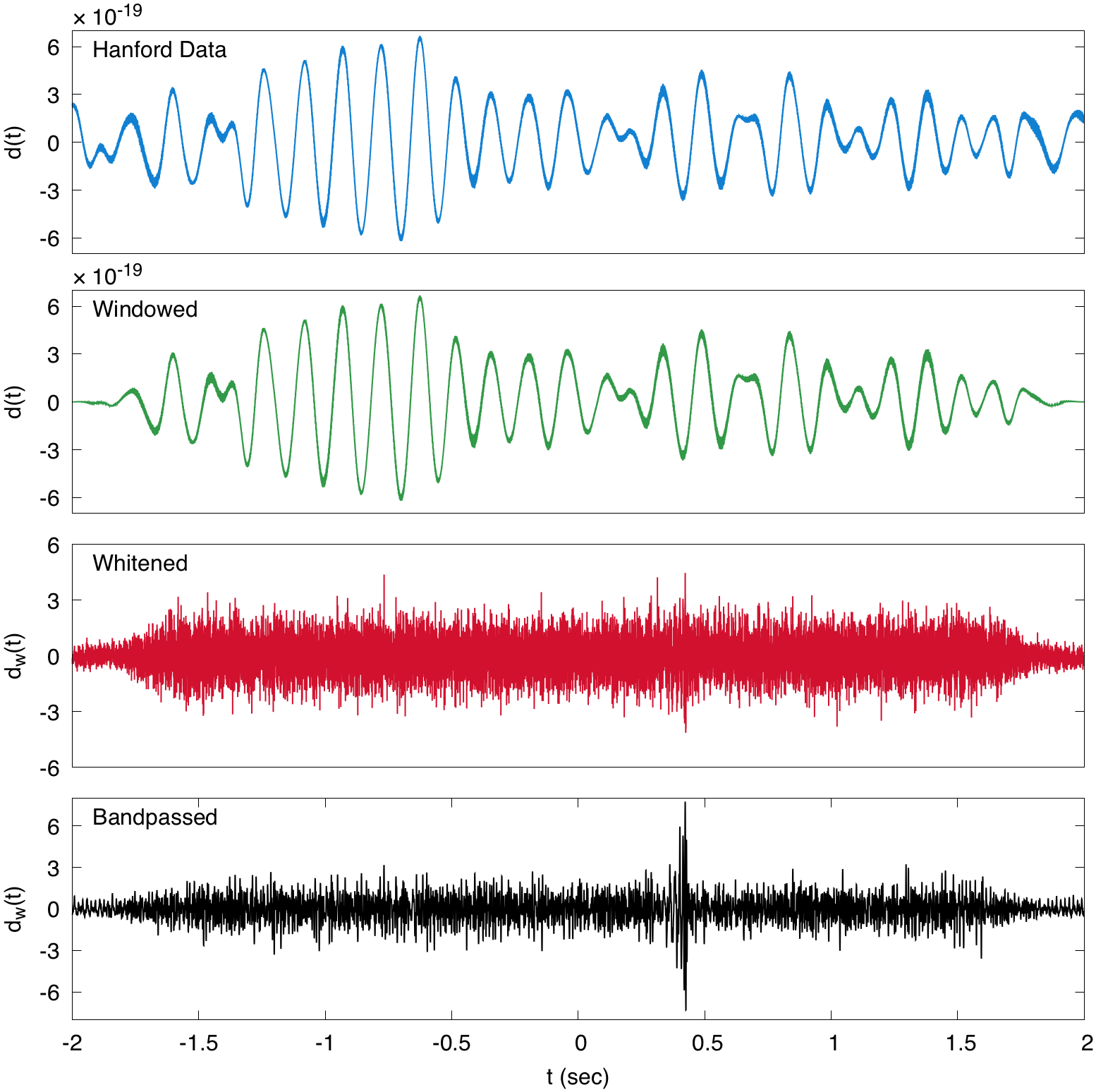}  
\caption{\label{fig:101}  A sequence of processing steps applied to the calibrated strain from the LIGO-Hanford detector showing 4 s of data centered on GPS time 1126259462 (September 14, 2015 09:50:45 UTC). First a Tukey window with 0.5 s roll-off is applied, then the data are whitened using an estimate of the noise spectral density. Finally
the data are bandpassed filtered to enhance features in the passband $[35 ~{\rm Hz}, 350 ~{\rm Hz}]$, revealing the presence of 
gravitational-wave signal GW150914.}
\end{figure}

As an illustration, Figure~\ref{fig:101} shows a sequence of processing steps applied to a stretch of calibrated strain data from the LIGO-Hanford detector around the time of GW150914. The raw data are dominated by low-frequency noise. A Tukey window with 0.5 s transition regions was applied to the raw data.  Next, the data were whitened by dividing the Fourier coefficients by an estimate of the amplitude spectral density of the noise, which ensures that the data in each frequency bin has equal significance by down-weighting frequencies where the noise is loud. The data were then inverse Fourier transformed to return to the time domain:
\begin{equation}
d(t) \;  \xrightarrow{{\rm FFT}}\;  \tilde d(f) \; \xrightarrow{{\rm Whiten}}\;  \tilde{d}_w(f) = \frac{\tilde d(f)}{S_n^{1/2}(f)}\;  \xrightarrow{{\rm iFFT}} \; d_w(t) \, .
\end{equation}
The whitened samples were scaled to have unit variance in the time domain. As a final step, the data were bandpass filtered using a zero-phase, eighth order Butterworth filter with pass band $[35 ~{\rm Hz}, 350 ~{\rm Hz}]$. The bandpass enhances the visibility of features of interest in this band by removing noise outside of the band -- seismic and related noise at low frequencies, and quantum sensing noise at high frequencies. 
Note that such narrow bandpassing is only used for visualization purposes and is not employed in the LVC analyses. The gravitational-wave signal GW150914 is visible in the whitened and bandpassed data shown in the lower panel of Figure~\ref{fig:101}.  

While the steps above can make loud transient signals like GW150914 more easily visible in the strain time series, LVC's statistical analysis pipelines typically use a different sequence of processing steps.
LVC pipelines for detection and parameter estimation proceed by first high-pass filtering the data, to remove high-amplitude noise below the range of frequencies that will be analyzed by the pipelines which typically starts at $\sim 20~{\rm Hz}$. The data may also be down-sampled, after low-pass filtering to avoid aliasing, to reduce computational costs; thus its frequency content will be affected by the anti-aliasing filter at high frequency, with a formal cutoff at the Nyquist frequency of the down-sampled data~\cite{Messick:2016aqy,Usman:2015kfa}.
The LVC parameter estimation pipelines do not apply any bandpass filter to the data, but limit the likelihood integral calculation to begin at some lower frequency cut-off (typically also $20~{\rm Hz}$). 

GW150914 was originally identified with high significance by a generic search for coherent excess power across the detector network~\cite{Abbott:2016blz,TheLIGOScientific:2016uux}, as well as by matched-filtering analyses~\cite{TheLIGOScientific:2016qqj}, as described in Section \ref{s:searches}, but this loud signal is also clearly visible in the data even with the minimal processing described here.

\subsection{Methods for measuring the noise spectrum}
The power spectral density of the noise $S_n(f)$ is not known {\it a priori} and must be estimated from the data. 
One can perform a complex FFT of the entire data stream around some time to be searched for signals, but that yields only two samples (real and imaginary parts) per frequency bin, hence the variance in the estimate of $S_n(f)$ in any single frequency bin is large.
To overcome this, either some form of averaging is used~\cite{Aasi:2013jjl}, or a fit is made to a physical model for the spectrum~\cite{Littenberg:2014oda}. For example, Welch averaging~\cite{1161901} can be used to reduce the variance in the estimated power spectrum, but at the cost of either reducing the frequency resolution or requiring longer 
stretches of data. The spectral estimate used to whiten the data in Figure~\ref{fig:101} was found by applying a Welch average to 1024 s of data centered on GPS time 1126259462 (the nearest integer GPS time to the peak of the GW150914 signal). The data were broken up into overlapping 4 s long chunks, each spaced by 2 s. The data in each chunk was Tukey filtered and Fourier transformed. The power spectrum from all the chunks was then averaged. 

\begin{figure}[h]
\centering \includegraphics[scale=0.4]{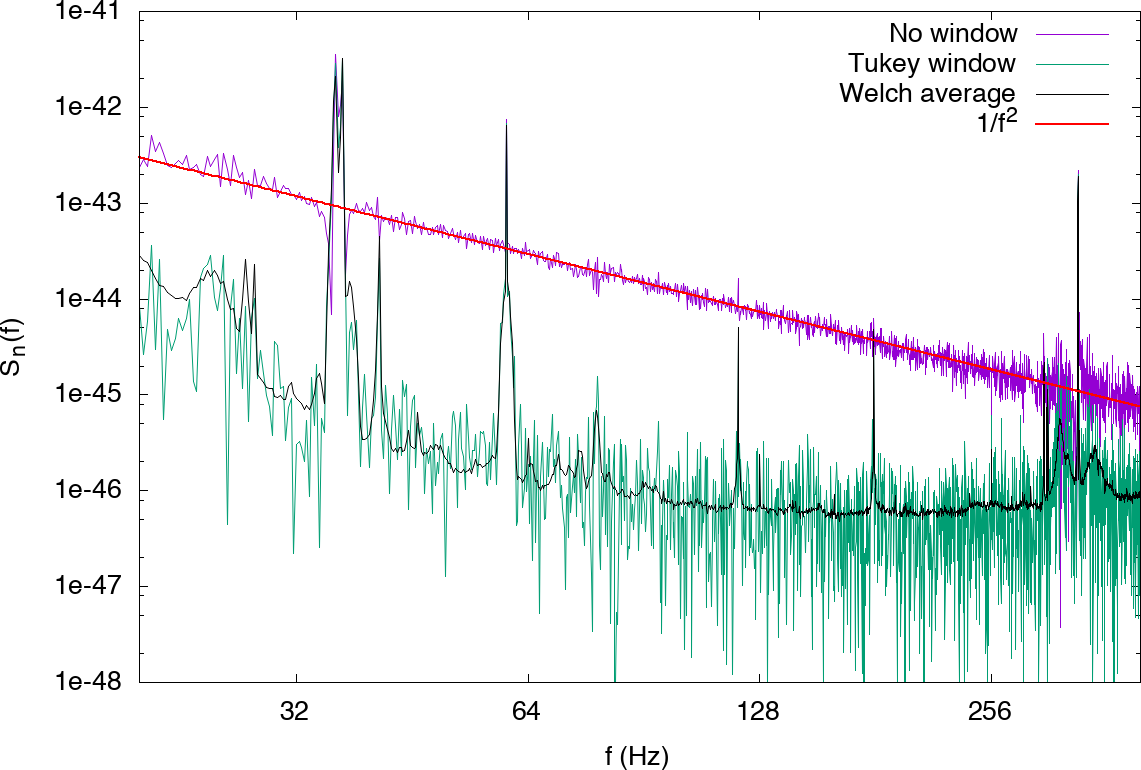}  
\caption{\label{fig:specH1}  Power spectral density for the data shown in Figure~\ref{fig:101}. The spectrum for the non-windowed data are swamped by spectral leakage, and follow a $1/f^2$ scaling. The Welch average was computed using a longer stretch of data.}
\end{figure}

Figure~\ref{fig:specH1} compares the power spectrum of the Hanford data shown in Figure~\ref{fig:101}, before and after applying the Tukey window, to the power spectrum estimated using Welch averaging.
The non-windowed spectrum is swamped by spectral leakage, and follows a $1/f^2$ scaling. This scaling results from the abrupt step function at the beginning and end of the data to be Fourier transformed. This non-windowed data chunk arises from multiplying a longer stretch of data by a boxcar (or top hat) window. Thus, when it is Fourier transformed, the result is the convolution of the desired spectrum of the original data with the Fourier transform of the 4s-long boxcar window, i.e.\ a cardinal sine (sinc) function whose amplitude decreases as $1/f^2$. Since the noise spectrum rises much more rapidly than $1/f^2$ towards low frequencies, the entire visible frequency range is then dominated by the leakage from this low-frequency component.


When the noise spectrum varies significantly over time
other spectral estimation methods have to be used~\cite{Allen:2005fk,Usman:2015kfa}. One approach used in LVC parameter estimation studies is to fit a parametrized spectral model to the data that has a smooth spline component and a collection of Lorentzian lines~\cite{Littenberg:2014oda}. 
In Section~\ref{sec:timefreq} we also discuss in detail the issue of stationarity and non-stationarity of the data, and the effects this has on the data analysis.

\begin{figure}[h]
\centering \includegraphics[scale=0.4]{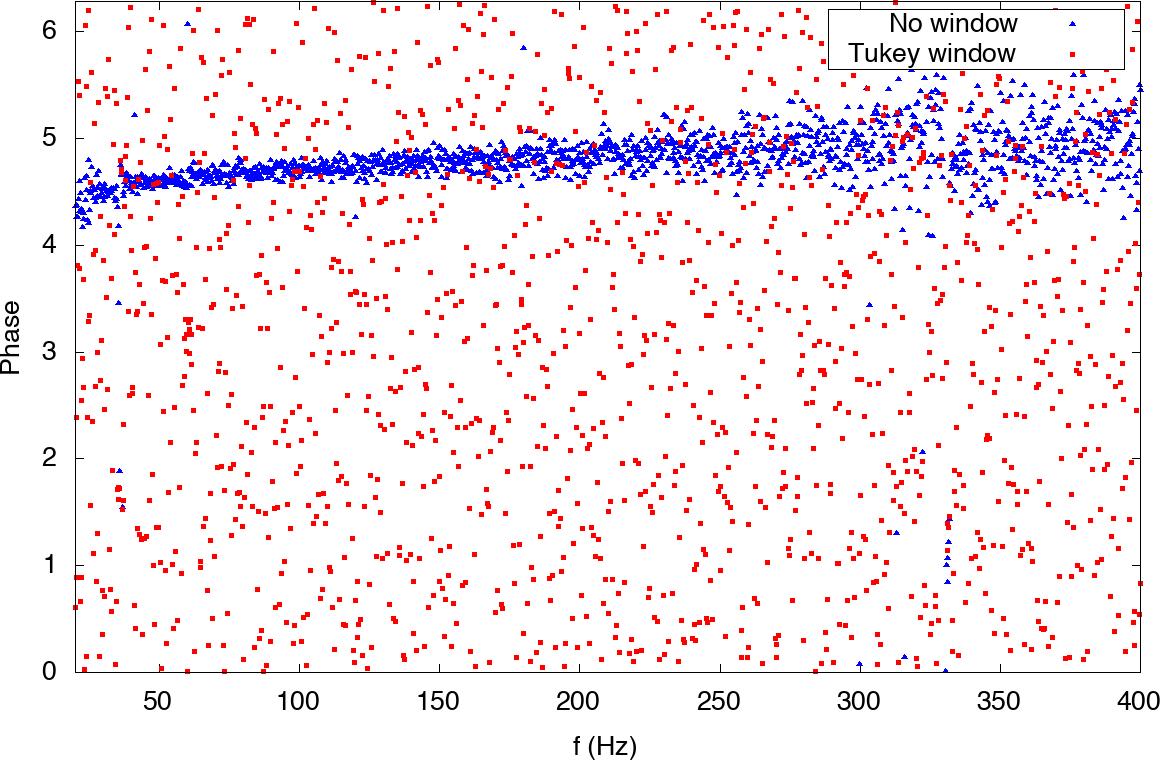}  
\caption{\label{fig:phaseH1}  The Fourier phases of the stretch of LIGO-Hanford data shown in Figure~\ref{fig:101}. If no window is applied before performing the FTT, as was the case in the analysis in~\cite{Creswell:2017rbh}, spectral leakage causes the phase to be correlated. When the Tukey window is applied the phases appear randomly distributed, as expected for Gaussian noise. The phases show some clustering around the $60~{\rm Hz}$ power line, consistent with the deterministic origin of this noise component.}
\end{figure}

In addition to causing spectral leakage, improper windowing of the data can result in spurious phase correlations in the Fourier transform. Figure~\ref{fig:phaseH1} shows a scatter plot of the Fourier phase as a function of frequency for the same stretch of data shown in Figure~\ref{fig:specH1}, both with and without the application of a window function. The un-windowed data shows a strong phase correlation, while the windowed data does not. 

\begin{figure}[h]
\centering \includegraphics[height=2.4in]{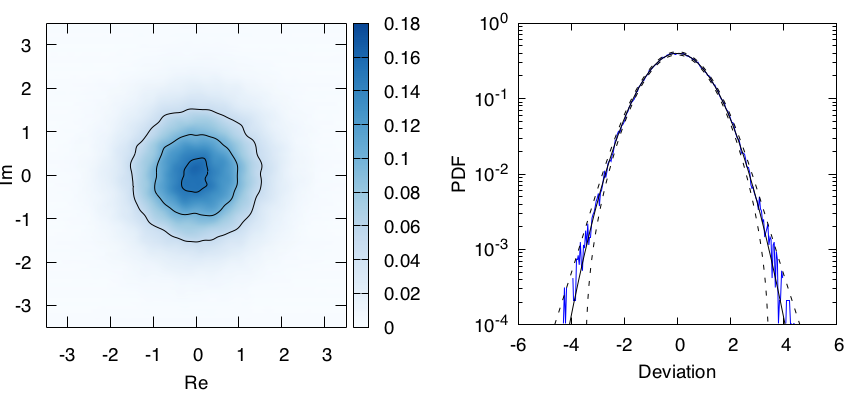}
\caption{\label{fig:clean} The panel on the left shows a 2-d density plot of the whitened real and imaginary Fourier amplitude deviations using 256 s of LIGO-Livingston data centered on GPS time 1186741733 covering the band from $32~{\rm Hz}$ to $512~{\rm Hz}$. The panel on the right shows a 1-d histogram of the Fourier amplitudes. The solid line is for a reference ${\cal N}(0,1)$ distribution, while the dashed lines indicate the expected 3-sigma variance from having a finite number of samples. }
\end{figure}

The degree to which a time series is consistent with being stationary and Gaussian noise can be diagnosed by looking at the distribution of its Fourier transformed frequency samples.
If the noise is stationary and Gaussian the real and imaginary parts of the whitened noise in each frequency bin will be a collection of independent and identically distributed (i.i.d.) random variables with zero mean and unit variance: $x \sim {\cal N}(0,1)$.
Departures from stationarity result in correlations between samples in different Fourier bins, while departures from Gaussianity can be identified by comparing the distribution of samples to a unit normal distribution. Loud instrumental noise transients and loud gravitational-wave bursts do contribute to non-stationary and non-Gaussian features, but away from these transient disturbances the LIGO-Virgo data can be approximated as stationary and Gaussian. Figure~\ref{fig:clean} shows the whitened Fourier amplitudes for a quiet stretch of data from the LIGO-Livingston observatory.

%% file: timefreq.tex
\section{Time-frequency analysis and stationarity}
\label{sec:timefreq}

The LIGO-Virgo data exhibit two main types of non-stationary behavior. The first is slow and continuous adiabatic drifts in the power spectrum occurring over minutes or hours, and the second is short-duration noise transients, which we refer to as \emph{glitches}, that are typically localized in time and frequency.
Additional non-stationarity has been observed in the vicinity of spectral lines, such as those due to electromagnetic couplings to the 50/60 Hz AC power supply. The adiabatic drifts in the power spectrum can be defined in terms of {\em locally stationary} processes~\cite{1057413,zbMATH00894812}.  A locally stationary process has a covariance function which is the product of a covariance function for a stationary process and a time-variable function.

\begin{figure}[h]
\centering \includegraphics[scale=0.9]{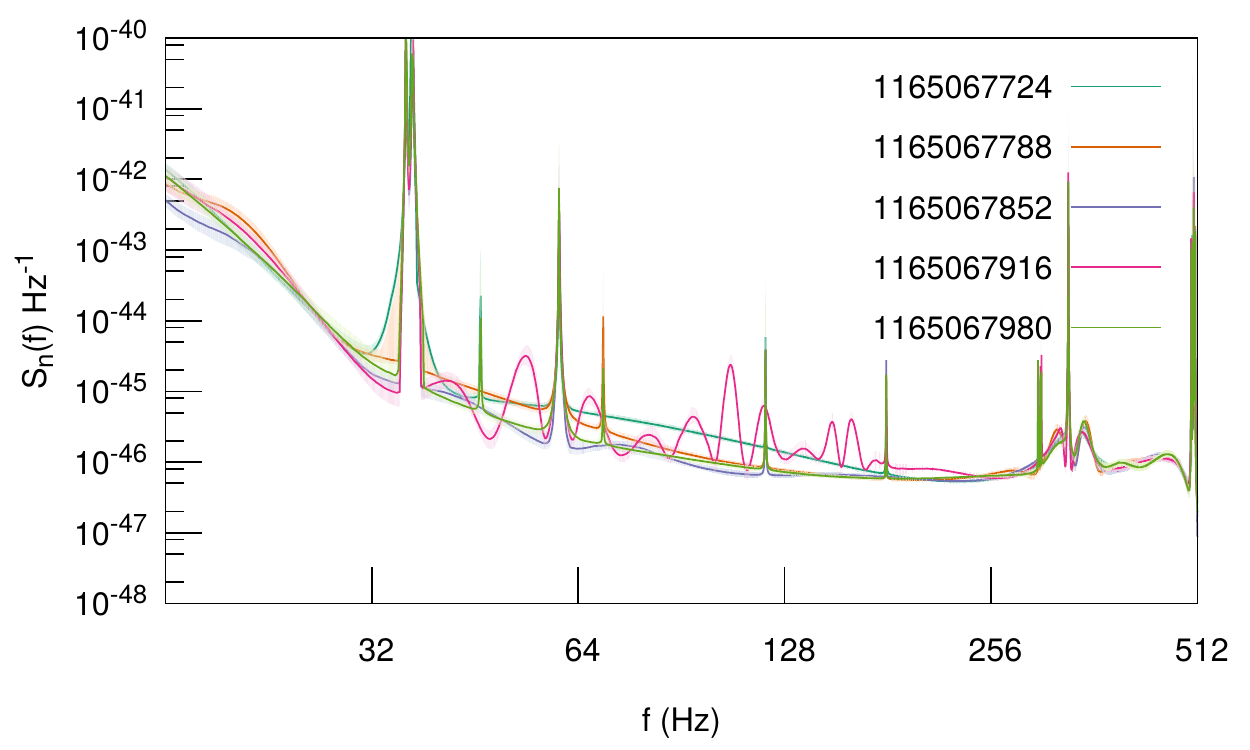} 
\caption{\label{fig:sft} Power spectral density (solid lines) with 90\% credible intervals (shaded bands) for the LIGO-Hanford detector computed using 8-second stretches of data spaced by 64-second intervals starting at GPS time 1165067724. During this time period there was significant broad-band non-stationarity between 32 and 256 Hz.}
\end{figure}

The stationarity of the data is evaluated as part of candidate event validation~\cite{TheLIGOScientific:2016zmo, LIGOScientific:2018mvr}.
Here we describe some simplified non-stationarity tests that can be applied to the data. 
Non-stationarity can in principle be identified by looking for correlations in the Fourier amplitudes, but it is easier to identify and classify non-stationary behavior using time-frequency methods. 
The simplest approach is to divide the data into small chunks of time centered on time $t_i$, and compute a smoothed estimate for the power spectrum for each chunk $S_n(f,t_i)$.  
Figure~\ref{fig:sft} shows Bayesian power spectral density estimates~\cite{Littenberg:2014oda} computed using 8-second stretches of data from the LIGO-Hanford instrument that are spaced at 64-second intervals.  
The instrument noise level was highly variable during this time period, showing large changes in the power spectral density in the band between 32 Hz and 256 Hz (note that this particular period of time for this example was chosen due to observed large variations in the detector's sensitivity).

\begin{figure}[h]
\centering \includegraphics[scale=0.35]{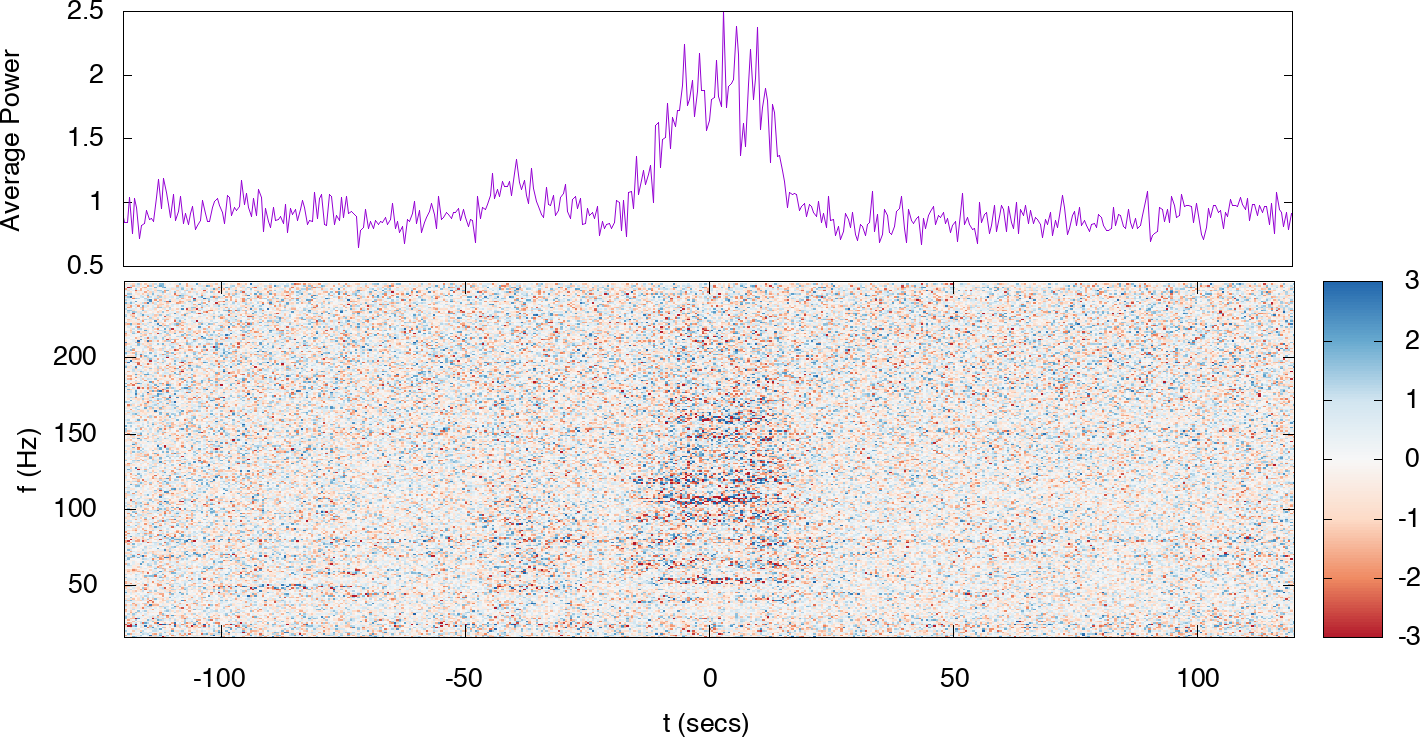} 
\caption{\label{fig:powtime} Fluctuations in the whitened data for the same stretch of highly non-stationary LIGO-Hanford data used to produce Figure~\ref{fig:sft}. The data were first whitened using an amplitude spectral density estimated from 256 seconds of data centered at GPS time 1165067917, 
then a discrete wavelet wavepacket transform
was used to produce the scalogram shown in the lower panel. The upper panel shows the average power as a function of time computed from the scalogram.}
\end{figure}

Wavelets provide a more flexible analysis framework than short-time Fourier transforms. Continuous wavelet transforms are commonly used in LIGO-Virgo data studies to produce spectrograms that provide a visual indication of non-stationary behavior. Quantitative assessments of non-stationarity may also be made by using discrete,  orthogonal wavelet transforms. These can be visualized using a \emph{scalogram}, showing the amplitudes of the wavelet basis functions at each discrete time and frequency pixel. Figure~\ref{fig:powtime} shows a scalogram of the same stretch of LIGO-Hanford data which were used to produce Figure~\ref{fig:sft}. The data were first whitened using an amplitude spectral density estimate taken from 256 seconds of data centered at GPS time 1165067917. The whitened data were then transformed using discrete wavepackets~\footnote{Note that the standard discrete wavelet transformation applies successive high and low pass filters in a particular order. The wavelet wave packet transform generalizes this to consider all possible combinations of filter application orders. A particular path through this transformation sequence defines some wavelet wave packet transformation. One is free to choose the decomposition path. Various criteria can be used to select an optimal or near optimal decomposition for a particular data set. In the study presented in this paper we use a path that gives a regular spacing in frequency bands because this choice provides a simple generalization of a discrete Fourier transform to the more flexible time-frequency case. For more information on this method, see~\cite{10.1117/12.55886}.}, 
built from Meyer wavelets~\cite{meyer1990ondelettes}, that were chosen to give uniform time and frequency coverage with tiles of size $\Delta t=0.5$ seconds and $\Delta f=1$ Hz. The average power at each time was then computed by summing the squares of the wavelet amplitudes (and dividing by a normalization constant) between 16 and 256 Hz. The noise level is elevated for almost a minute around the center of the data segment.

When this analysis is applied to stationary, Gaussian noise, the power in each time interval follows a chi-squared distribution with $N_f$ degrees of freedom, where $N_f$ is the number of frequency pixels that are summed over. 
The distribution of the average power can be compared to this reference distribution using e.g.\ an Anderson-Darling test~\cite{AndersonDarling}, to yield a quantitative measure of the non-stationarity. Note that while stationary noise is stationary no matter what time span is considered, non-stationary noise will produce different measures of departure depending on the averaging scale (here the width of the wavelet pixels in time) and time span of the data. For visualization purposes it is convenient to transform the average power $p(t)$ to a new variable $s(t)$ via the Wilson-Hilferty transformation~\cite{10.2307/86022}, 
such that $s(t)$ follows a ${\cal N}(0,1)$ Gaussian distribution when the noise is stationary and Gaussian. 

\begin{figure}[h]
\centering \includegraphics[scale=0.35]{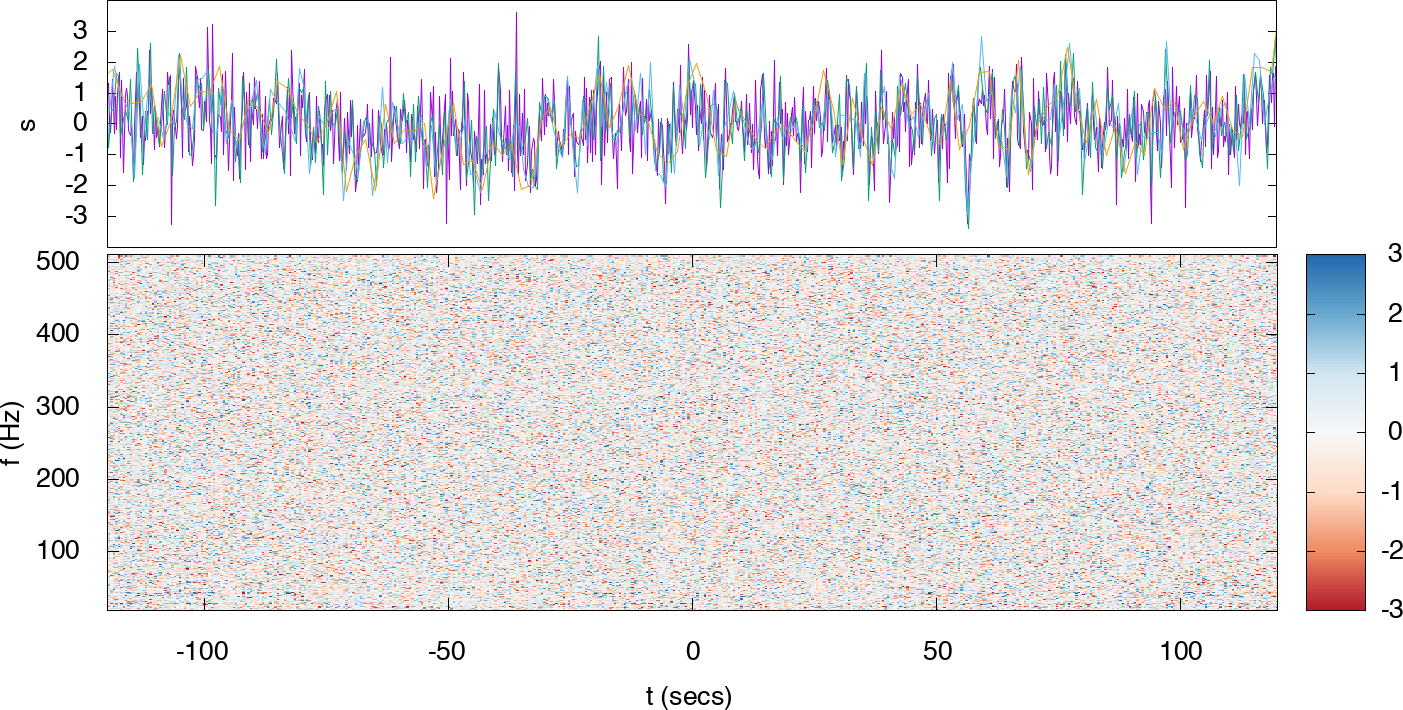} 
\caption{\label{fig:stat1}  A quiet stretch of whitened strain data from the LIGO-Livingston laboratory centered on GPS time 1186741733. The upper panel shows the transformed average power statistic $s(t)$ for a variety of wavelet resolutions (plotted in different colors) with pixels ranging from 0.25 seconds to 2 seconds in width. The power fluctuations $s(t)$ should follow a zero mean, unit variance Gaussian distribution when the noise is stationary and Gaussian. The lower panel shows a wavelet scalogram at 0.5 second resolution. }
\end{figure}

\begin{figure}[h]
\centering \includegraphics[scale=0.35]{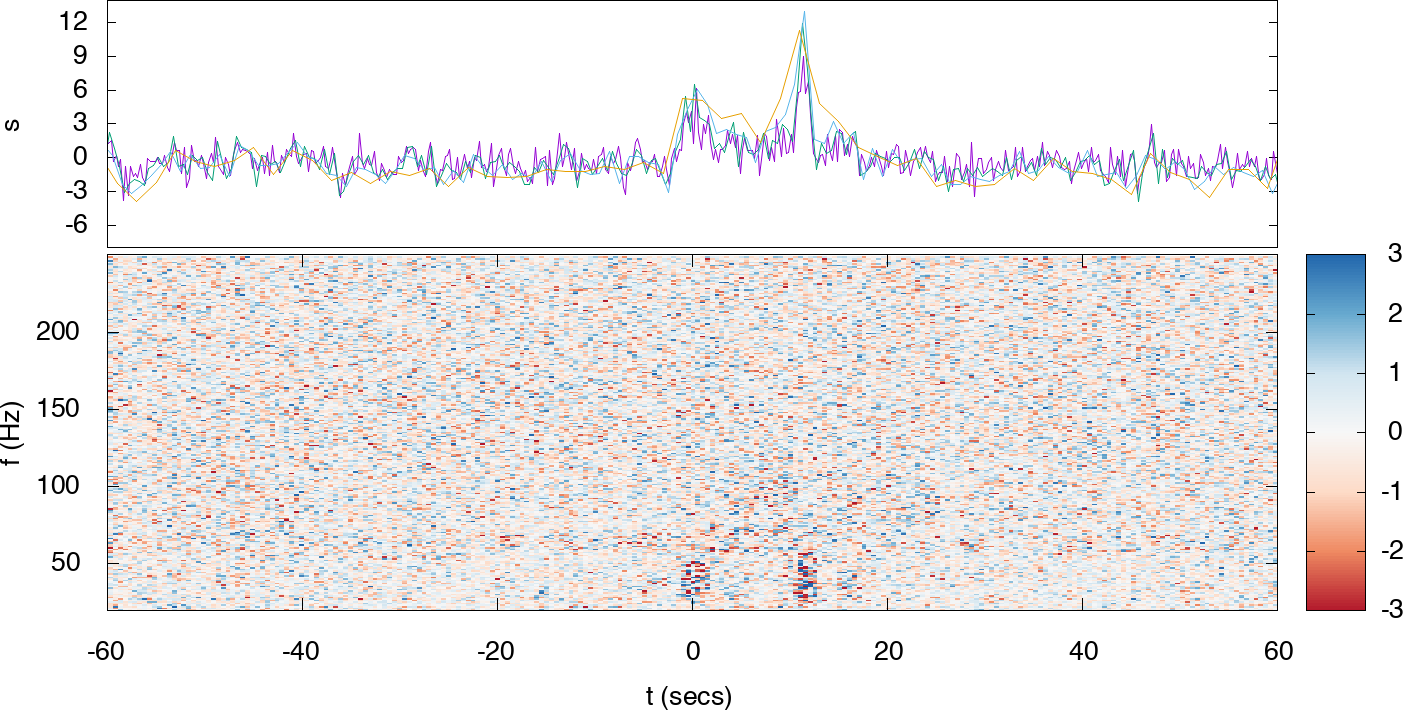} 
\caption{\label{fig:stat2}  A stretch of whitened strain data from the LIGO-Livingston laboratory centered on GPS time 1166358283. The upper panel shows the transformed average power statistic $s(t)$ for a variety of wavelet resolutions with pixels ranging from 0.25 seconds to 2 seconds in width. The power fluctuations $s(t)$ should follow a zero mean, unit variance Gaussian distribution when the noise is stationary and Gaussian. The lower panel shows a wavelet scalogram at 0.5 second resolution. A series of glitches causes significant non-stationarity.}
\end{figure}

Applying the Anderson-Darling test to the total power yields $p$-values for the hypothesis that the data are stationary. When applied to the quiet stretch of data shown in Figure~\ref{fig:stat1} the test yields a p-value of $p= 0.74$, indicating that the hypothesis that the data are stationary cannot be rejected over this time period at this wavelet scale. 
Applying the same test to the data shown in Figure~\ref{fig:stat2} yields a $p$-value of $p=2.3\times 10^{-6}$, and we can reject the hypothesis that the data are stationary with high confidence.
Any analysis that attempted to detect or estimate the parameters of a possible gravitational-wave signal occurring in this stretch of data would then have to take steps to mitigate, suppress or otherwise account for the departure from stationary noise.

%% file: further-reading.tex
\section{Detector calibration and data quality}\label{s:further-reading}

In this section we provide the central concepts related to the calibration of the data as well as an overview of the data quality checks we perform. These procedures ensure that the strain data used for analyses (namely, the analyses used by the LVC in publication results) and made public on the GWOSC is calibrated properly with known error bars, and that time periods of poor data quality can be avoided, as explained below.

\subsection{Detector calibration} \label{ss:calibration}

\begin{figure}[t]
\includegraphics[width=\textwidth]{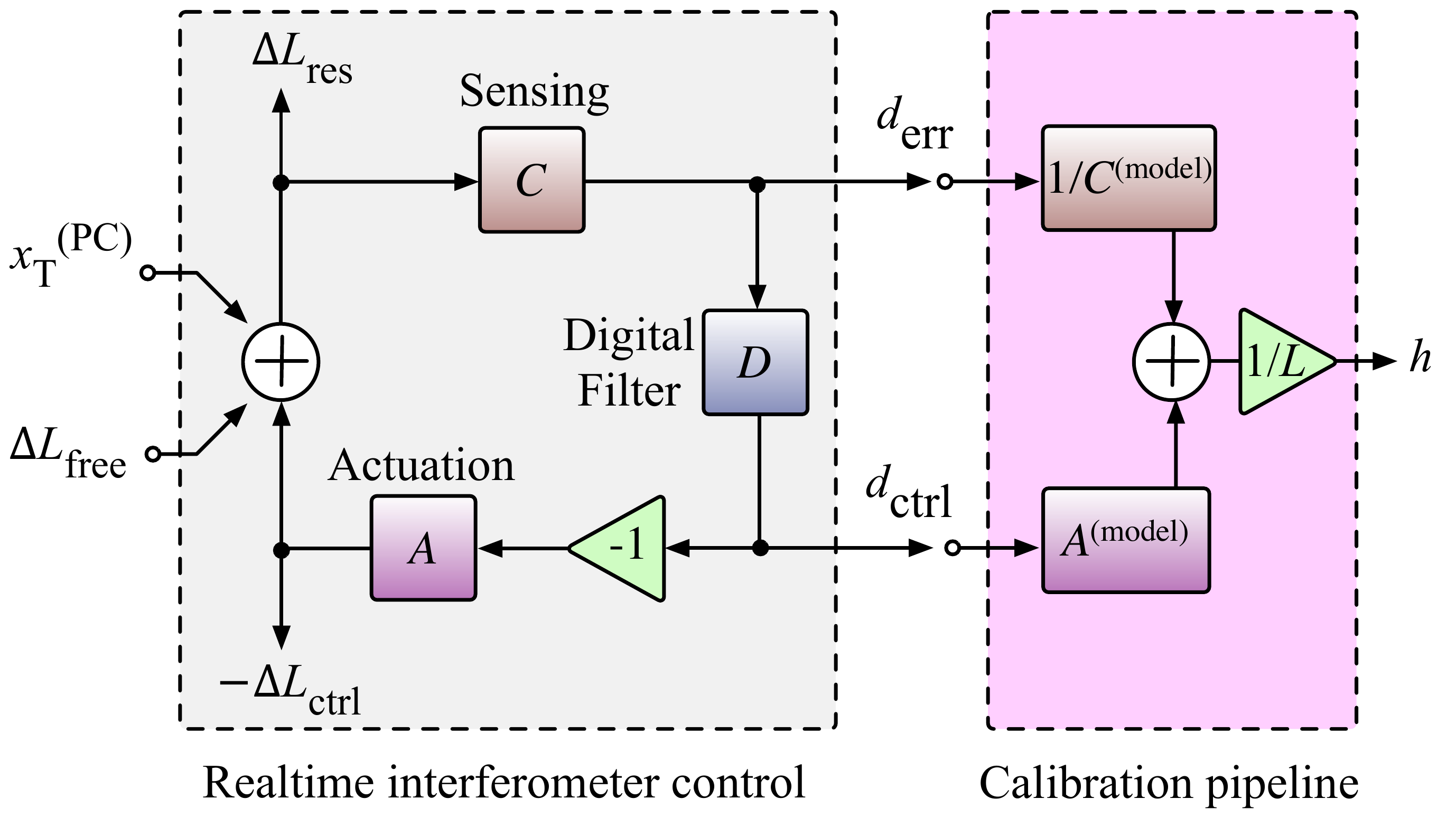}
\caption{Differential arm length control loop and calibration diagram of the LIGO detectors from the GW150914 companion paper on calibration \cite{TheLIGOScientific:2016agk}. The left (grey) box shows the realtime detector controls while the right (purple) box shows the calibration procedure. $\Delta L_{free}$ is the unsuppressed change in differential arm length, and hence the desired quantity.
The photodiodes (part of the sensing $C$) measure the residual differential arm length $\Delta L_{res}$, which is suppressed by the feedback loop.
The ``error signal'' $d_{err}$, equal to $\Delta L_{res}$ multiplied by the sensing function $C$, 
is passed through digital filters, $D$, and applied to the differential arm length actuators through the actuation function, $A$. In order to reconstruct an estimate of $\Delta L_{free}$ in units of strain we model $A$ and $C$, denoted in the purple box. $x_{T}^{(PC)}$ denotes where in the loop we apply a force to the test mass mirrors, via radiation pressure (photon calibrator), in order to measure $A$ and $C$, as functions of frequency. The output of the calibration pipeline is then a strain signal, $h(t)$, that is a faithful representation of $\Delta L_{free}/L$.}
\label{f:cal}
\end{figure}

The Advanced LIGO~\cite{TheLIGOScientific:2014jea,Martynov:2016fzi,TheLIGOScientific:2016agk,Abbott:2016jsd} and Advanced Virgo~\cite{TheVirgo:2014hva,Acernese:2018bfl,casanuevadiaz:tel-01625376} detectors use feedback loops to keep the optical cavities on resonance. 
The strain calibration must thus include models and measurements of all readout electronics, as well as of electronics and transfer functions of actuation hardware that act on the mirrors through multiple points in the suspension systems \cite{Robertson_2002}. As shown in Figure~\ref{f:cal}, there are three main components of the differential arm control loop for Advanced LIGO: the actuation function $A(f)$, the sensing function $C(f)$, and the digital filters applied, $D(f)$. All three are measured and modeled as functions of frequency.
 
The digital filters are known to great precision so the calibration error and uncertainty come from the differences between the model and measurement (including measurement error) of the actuation and sensing functions, $A$ and $C$. To independently measure the actuation and sensing functions, a pair of beams from auxiliary lasers are reflected off of each test mass mirror, with their intensities modulated at a known frequency and amplitude to actuate with radiation pressure.  These auxiliary laser assemblies are referred to as photon calibrators~\cite{Karki2016}. Once $A$ and $C$ are known, the true differential arm length is extracted and translated to a strain by dividing by the common length of the arm $L$ (4~km for LIGO, 3~km for Virgo), per the following equation:
\begin{equation}\label{calib}
h(t) =  \frac{1}{L}\left[ \mathcal{C}^{-1}*d_{err}(t)+\mathcal{A}*d_{ctrl}(t)\right] ,
\end{equation}
where $\mathcal{C}$ and $\mathcal{A}$ are time-domain filters derived from frequency-domain measurements of the actuation function $A(f)$ and the sensing function $C(f)$.

Note that the gravitational-wave strain can also appear in the common-mode arm length changes, and in changes to the lengths of all degrees of freedom in the detector. However, only the sensing of the differential change in the interferometer's arm lengths is engineered to have low enough instrumental noise to be sensitive to the strain induced by gravitational waves. The other optical lengths are controlled in order to maintain an optimal and linear response to the gravitational-wave strain in the differential degree of freedom.

Calibration measurements are made periodically in each observing run. In addition, to monitor time dependent parameters such as optical gain, cavity pole frequency and actuation strength drifts, several \emph{calibration lines} are continously injected, at specific frequencies, by applying sinusoidal forces on the test mass mirrors using the photon calibrators; these lines will be present in the raw strain data. The calibration line frequencies are different amongst the detectors. 

For the second observing run O2 and onwards, the calibration lines are removed from the calibrated $h(t)$ strain data channel (within the calibration accuracy)~\cite{Driggers:2018gii,Acernese:2018bfl}; the calibration lines were not removed from the O1 data~\cite{Abbott:2016jsd}.
Even for O1, the presence of the calibration lines does not affect the search for compact binary coalescence gravitational-wave signals as the amplitude of data at the frequencies of the lines is suppressed via the whitening~\cite{Cuoco:2000gv,Cuoco:2001wz} of the data when the calculation of the detection statistic is made for the data from each detector (see Section~\ref{s:searches}).
Similarly, for parameter estimation the presence of the noise spectral density in the likelihood (see Section~\ref{s:pe}) minimizes the influence of spectral lines including calibration lines.  Because the spectral lines are narrow, this frequency-domain weighting has a negligible effect on signal searches and parameter estimation and does not lead to any spurious effects such as generation of false candidate events or parameter biases.

For the Advanced Virgo calibration in O2 it was necessary to account for the transfer function of the optical response of the interferometer. This requires a calibration of the longitudinal actuators for the mirrors, still based on the laser wavelength as length reference using the so-called {\it free swinging Michelson} configuration described in~\cite{Accadia:2010aa}. In addition, the interferometer's output power as determined by the readout electronics also requires calibration.  Calibration measurements are made weekly in each observing run and have shown stable actuation strengths. To monitor the time dependent optical gain and cavity pole frequency, several calibration lines are continously injected, as in LIGO detectors, at specific frequencies, by applying sinusoidal forces on the test mass mirrors using the electro-magnetic actuators.  By construction, the calibration lines are removed from the calibrated $h(t)$ strain data channel (within the calibration accuracy).
For Advanced Virgo in O2 the gravitational wave strain reconstruction removed the contributions from control signals to the test mass mirror motion.
A {\it photon calibrator}, namely an auxiliary laser that is used to reflect photons off a mirror and induce momentum transfer, was used to verify the calibration and confirm the sign of the strain channel $h(t)$. See~\cite{Acernese:2018bfl} for more explicit details of the Advanced Virgo calibration system for O2.

Calibrated  strain  data for the LIGO and Virgo detectors are created online for use in low-latency searches.  After the completion of the observing runs, final time-dependent calibrations were generated for each detector. The results presented in GWTC-1~\cite{LIGOScientific:2018mvr} use the full frequency-dependent calibration uncertainties described in~\cite{Cahillane:2017,Viets_2018,Acernese:2018bfl}. 
It is important to note that the detector strain channel $h(t)$ is only calibrated between 10 Hz and 5 kHz for Advanced LIGO and 10 Hz and 8 kHz for Advanced Virgo~\cite{Acernese:2018bfl}; the channel is not a faithful representation of strain at lower or higher frequencies.

\subsection{Data quality and terrestrial noise}\label{ss:DQ-Terr-Noise}

As described in Sections \ref{sec:noise} and \ref{sec:timefreq}, calibrated LIGO and Virgo data can be both non-stationary and non-Gaussian at certain times and frequencies.  Glitches may mimic true transient astrophysical signals in individual detectors \cite{TheLIGOScientific:2016zmo}, while spectral lines such as those seen in Figure \ref{f:asds} can blind searches for long-duration signals at those specific frequencies \cite{Covas:2018}.  In this section we outline how we identify and characterize these noise features so that we can either exclude the bad data or assess the impact of remaining artifacts on searches for gravitational-wave signals.

\begin{figure}[t]
\center
\includegraphics[width=0.5\textwidth]{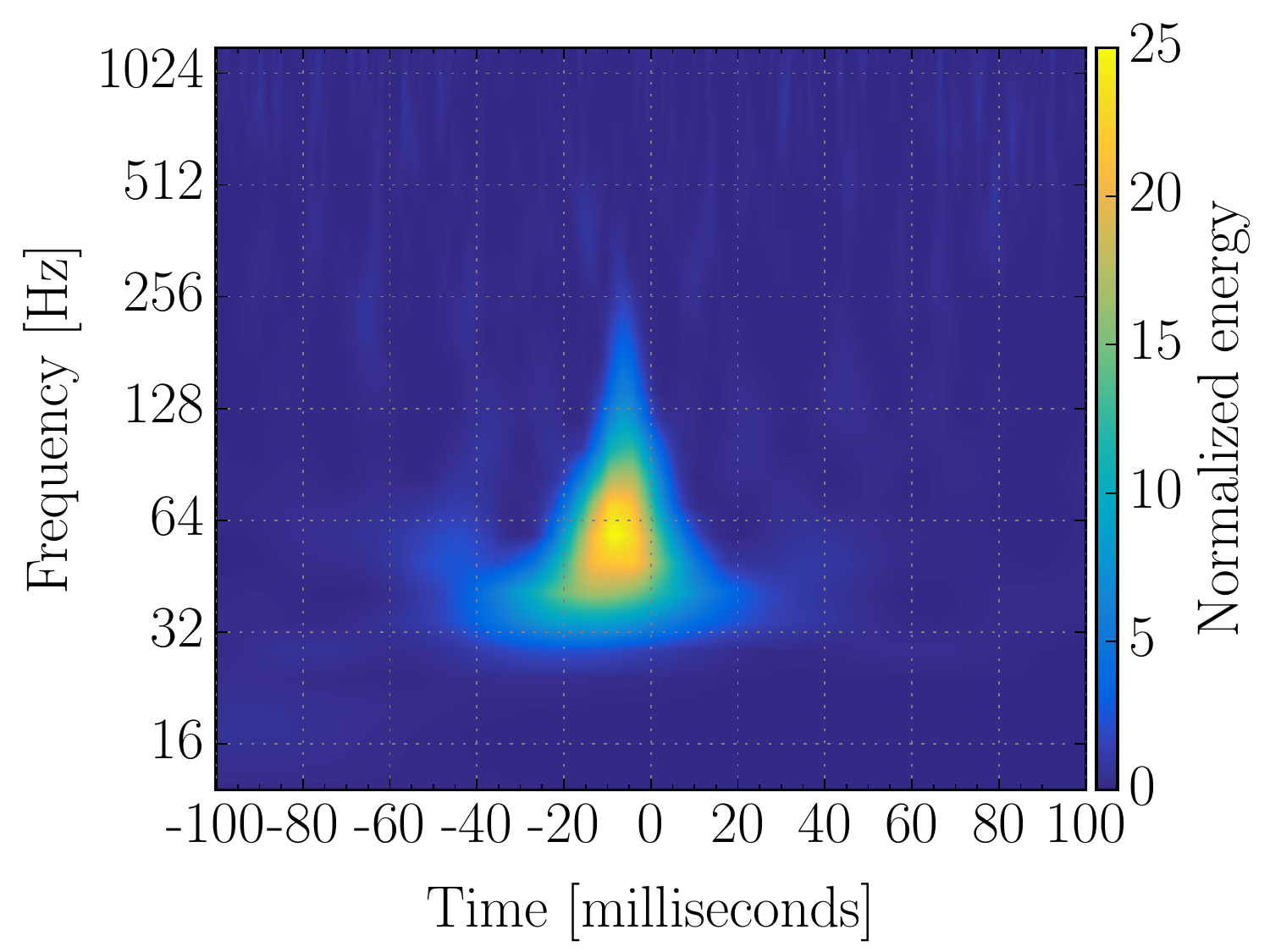}
\caption{Spectrogram of an example of transient noise in the advanced detectors; a \textit{blip} glitch in LIGO-Livingston.  This is a zoomed image of the blip glitch shown in Fig.\ 10 of \cite{TheLIGOScientific:2017lwt}.}
\label{f:glitch}
\end{figure}

Figure \ref{f:glitch} shows an example of a glitch. Glitches with power comparable to detectable signals have historically occurred on the order of once per minute, with larger glitches occurring less frequently. Even in their nominal state, the detectors' data contain glitches
introduced by behavior of the instruments or complex interactions between the instruments
and their environment.
Many of these glitches (but not all) can be associated with transient signals in auxiliary channels from various sensors which serve as ``witnesses'' to environmental disturbances coupling into the interferometer.  These associations allow us to identify and catalog certain classes of glitches.
See~\cite{TheLIGOScientific:2016zmo} for a detailed presentation on the characterization of transient noise in Advanced LIGO, especially pertaining to the observation of the gravitational-wave signal GW150914.
Descriptions of various noise sources for Advanced Virgo in O2 can be found in~\cite{Cirone:2019zwq,Cirone:2018vdc,Bonnand_2017,ACERNESE2020102386}.


In searches for transient gravitational-wave signals, identified glitches and periods of poor data quality are flagged~\cite{TheLIGOScientific:2017lwt,Aasi:2014mqd,Aasi:2012wd}. Periods of data are vetoed at various levels or categories depending on the severity of the problems; the GWOSC open data releases make this information available~\cite{LOSC}. Sections of strongly non-stationary data that would corrupt the noise power spectral density estimates are removed entirely from the searches. Times when noise sources with known physical coupling to the gravitational-wave strain channel of the detector are active, and thus likely to cause glitches, are identified, and candidates at or around these times may be removed (vetoed) from search results. For a more detailed explanation of the strategy for mitigating noise sources, see Section 4 of~\cite{TheLIGOScientific:2016zmo}.
For searches for long-duration signals, frequency bands known to be dominated by instrument noise are omitted from the analysis~\cite{Covas:2018}. 
The strategy of vetoing times of probable glitches is expected to increase confidence in detection candidates that survive the application of vetoes~\cite{TheLIGOScientific:2017lwt,Isogai_2010,Smith_2011}, and may thus increase the number of astrophysical signals that can be confidently detected.  Detection methods are discussed in detail in Section~\ref{s:searches}.

Although the great majority of transient noise sources are of local origin and thus uncorrelated between detectors, some noise sources exist that are potentially correlated between detectors, such as electromagnetic pulses from lightning coupling inductively into the detectors~\cite{0264-9381-34-7-074002}.
A key feature of the LIGO and Virgo experimental design is an 
array of physical environment monitors designed to detect environmental disturbances, 
and to have greater sensitivity to those disturbances than the detector's gravitational-wave strain channel does.
The LIGO environmental sensor array includes seismometers, microphones, accelerometers, radio receivers and magnetometers to monitor ambient noise~\cite{Effler_2015} \footnote{See also http://pem.ligo.org}. Virgo has a similar array of sensors~\cite{TheVirgo:2014hva} \footnote{See also https://tds.virgo-gw.eu/?content=3\&r=15647}.
The environmental sensors' sensitivities are verified via a suite of noise injections performed at the beginning and end of each observing run; acoustic, magnetic, radio frequency, and vibrational tests are done to quantify the coupling from ambient noise to the gravitational-wave strain data $h(t)$~\cite{TheLIGOScientific:2016zmo}. These injections are conducted at multiple locations around the detector such that sensor coupling functions to $h(t)$ via multiple potential coupling paths are verified and well understood.
 

As shown in Figure 2 of \cite{TheLIGOScientific:2016zmo}, the external transient electromagnetic coupling of the ambient noise to the gravitational-wave data channel $h(t)$ is on the order of a factor of 100 below the current strain level, such that any electromagnetic source would have to register in one of the magnetometers surrounding the detector an SNR of 100 before registering in the gravitational-wave signal channel. This is easily confirmed with the study of lightning strikes during nearby storms~\cite{TheLIGOScientific:2016zmo}. The description of the coherence between the detectors' output strain signal $h(t)$ and magnetometers about the detector for the AC power frequencies (50/60 Hz) is nontrivial and described in \cite{DavisMassinger2018}. 
In Virgo, a detailed study of the electromagnetic coupling to the gravitational-wave data channel was recently carried out~\cite{Cirone:2018vdc,Cirone:2019zwq}.

 
In addition, a potential correlated noise source in searches for a stochastic gravitational-wave background is Schumann resonances, or low-frequency magnetic field resonances between the Earth's surface and the ionosphere excited by lightning~\cite{PhysRevD.87.123009,PhysRevD.90.023013}. These resonances are also monitored with sensitive magnetometers, with the future goal of subtracting their effect on gravitational-wave strain data \cite{Coughlin:2018str}. 
The effect of Schumann resonances on the measured gravitational-wave strain is below the current Advanced LIGO - Advanced Virgo noise floor~\cite{LIGOScientific:2019vic}.

%% file: likelihood.tex
\section{Noise model and likelihood}
\label{sec:likelihood}

The likelihood that the gravitational-wave strain data contains a given signal is the central quantity in both detection and parameter estimation of gravitational-wave events. 
In this section we relate this likelihood to the model assumptions commonly made for the noise components of gravitational-wave detector strain data.
The data time series ${\bf d}$ collected from an interferometer can be written as the sum of the gravitational wave response of the
detector, $\mathfrak{h}$, and the combination of all the noise sources in that detector ${\bf n}$ such that ${\bf d}={\bf n}+\mathfrak{h}$. 
Since the true gravitational-wave signal in the detector $\mathfrak{h}$ is unknown, we resort to using signal models denoted by $\bf{h}$. 
We consider a model $\bf{h}$ to be a good description of the signal in the data if the residuals ${\bf r} = {\bf d} - \bf{h}$ are consistent with our model for the instrument noise. 
More quantitatively, the likelihood that the data ${\bf d}$ contains a possible signal $\mathfrak{h}$ is given by the probability that ${\bf r}$ is a realization of the noise model. 
In other words, {\em the likelihood function is the noise model}. For Gaussian noise the likelihood can be written:
\begin{equation}\label{like}
p({\bf d} | {\bf h}) =  \frac{1}{ {\rm det}(2 \pi {\bf C})^{1/2}} \, e^{- \frac{1}{2} \chi^2({\bf d} , {\bf h})}\, ,
\end{equation}
where ${\bf C}$ is the noise correlation matrix, and 
\begin{equation}\label{chi}
 \chi^2({\bf d} , {\bf h}) = {\bf r} \cdot {\bf C}^{-1} \cdot {\bf r}  = (d_{Ik} - h_{Ik}) C^{-1}_{(I k)(J m)} (d_{Jm} - h_{Jm}) \, .
\end{equation}
The repeated indices include a sum over the network of detectors $I, J$ and the data samples $k$ and $m$. If the noise is uncorrelated between detectors $C_{(Ik)(Jm)} = \delta_{IJ} S^I_{km}$, where $S$ is the noise spectral density. Moreover, if the noise is stationary -- so that correlations depend on only the time lag between data samples -- then the noise correlation matrix in each detector will be diagonal in the Fourier domain: $S^{I}_{km} = \delta_{km} S^{I}(f_k)$. In that case we have $\chi^2({\bf d} , {\bf h}) = ({\bf r} | {\bf r})$ where $({\bf a} | {\bf b})$ is the familiar noise-weighted inner product:
\begin{equation}\label{nwip}
({\bf a} | {\bf b})  =2 \int_0^\infty \frac{ \tilde{a}(f) \tilde{b}^*(f) + \tilde{a}^*(f) \tilde{b}(f) }{S_n(f)} \, df \, .
\end{equation}
The likelihood function (\ref{like}) is central to Bayesian inference~\cite{Gregory:2005:BLD:1051497}, and with the specification of priors~\cite{box2011bayesian} for the signal and noise models, allows for the calculation of the model evidence~\cite{skilling2006,Veitch:2009hd} -- giving the odds that a signal is present -- and posterior distributions for the model parameters, $\boldsymbol{\theta}$, such as the masses and spins of a binary system~\cite{PhysRevD.64.022001,vanderSluys:2007st,Veitch:2014wba}.  For stationary, Gaussian noise that is uncorrelated between detectors, the likelihood function takes the form
\begin{equation}\label{like2}
p({\bf d} | \boldsymbol{\theta}) =  \exp\left(- \frac{1}{2} \sum_{I} \left[ ({\bf d}_I - {\bf h}_I(\boldsymbol{\theta}) | {\bf d}_I - {\bf h}_I(\boldsymbol{\theta} )) + \int \ln(S^{I}_{n}(f)) \, df \right]\right) \, ,
\end{equation}
where the sum is taken over the detectors in the network and $(a|b)$ is the noise weighted inner product defined in equation (\ref{nwip}). 

The likelihood function can also be used to define a frequentist detection statistic~\cite{PhysRevD.46.5236,creighton2012gravitational} given by the likelihood ratio between a signal $\bf{h}$ being present or absent in the data. If the data were stationary and Gaussian this statistic would follow a known distribution and the false alarm rate for an event could be computed analytically. In practice the noise exhibits deviations from stationarity and Gaussianity, and the methods used to detect and characterize signals have to be modified. Robust search methods have been developed that take into account the measured properties of the noise. These are described in Section~\ref{s:searches}. The noise modeling and consistency checks applied to signal characterization and parameters estimation are described in Section~\ref{s:pe}.

%% file: search.tex
\section{Signal detection}
\label{s:searches}

In this section we describe how candidate signals are identified in LIGO-Virgo data and how the statistical significance of each candidate is quantified by comparison with the observed properties of the detector noise.

\subsection{Model comparison and the matched filter}
\label{ss:matchedfilter}

The LVC searches for gravitational waves compare the null hypothesis, ${\cal
H}_0$, that a given stretch of data contains only noise, to the signal hypothesis, ${\cal
H}_1$, that the stretch of data contains both noise and a gravitational-wave signal\footnote{In reality all LIGO-Virgo data may contain some level of gravitational-wave signal, but a signal can only be detected if the null hypothesis is sufficiently disfavored relative to the signal hypothesis.}. Most searches
assume that general relativity correctly describes the gravitational-wave signals.
The likelihood of observing the data under the two hypotheses can be written
in terms of Eq.~(\ref{like}) by
\begin{equation}
p({\bf d} \mid {\cal H}_0) = p_{0}({\bf d})  \quad \text{and} \quad
p({\bf d} \mid {\cal H}_1) = p_{1}({\bf d})
\end{equation}
where ${\cal H}_0$ assumes noise alone with no signal in the data,
while ${\cal H}_1$ assumes a signal parameterized by $\boldsymbol\theta$,
${\bf h}({\boldsymbol\theta})$ is present in addition to the noise.
For the present, we will assume that each detector data stream is being analyzed
independently, and we discuss below how data from multiple detectors is
combined in LVC searches.
The probability of the signal hypothesis given
the observed data, known as the \emph{posterior probability}, is given by Bayes' theorem as
\begin{equation}
p({\cal H}_1 \mid {\bf d}) =
\frac{p({\cal H}_1) p_{1}({\bf d})}%
{p({\cal H}_0) p_{0}({\bf d})
+ p({\cal H}_1) p_{1}({\bf d})}
= 
\frac{p_{1}({\bf d})}{p_{0}({\bf d})}
\left[
\frac{p_{1}({\bf d})}{p_{0}({\bf d})}
+
\frac{p({\cal H}_0)}{p({\cal H}_1)}
\right]^{-1}
\end{equation}
where $p({\cal H}_0)$ and $p({\cal H}_1)$ are our prior beliefs of whether
a signal is absent or present in the data.  Regardless of these prior
beliefs, the posterior probability is
seen to be monotonic in the \emph{likelihood ratio}
\begin{equation}
   \Lambda({\bf d}|\boldsymbol{\theta}) = \frac{p({\bf d} \mid {\cal H}_1)}{p({\bf d} \mid {\cal H}_0)} = \frac{p_{1}({\bf d})}{p_{0}({\bf d})}
\end{equation}
and so this quantity is the optimal test statistic~\cite{Jaranowski2012}.
For Gaussian noise, the log of the likelihood ratio can be written in terms
of the inner product of Eq.~(\ref{nwip}) as
\begin{equation}\label{e:lnlambda}
   \log \Lambda({\bf d}|\boldsymbol{\theta})  =  ({\bf d} \mid {\bf h}(\boldsymbol{\theta}))-\frac{1}{2}({\bf h}(\boldsymbol{\theta}) \mid {\bf h}(\boldsymbol{\theta}))  \, .
\end{equation}
Only the first term of this expression involves the data; it is then observed
that the posterior probability is a monotonic function of
$({\bf d} \mid {\bf h}(\boldsymbol{\theta}))$, a quantity known as the
\emph{matched filter}, which, therefore, is also an optimal test statistic.

\subsection{Signal-to-noise ratio and template banks}
\label{ss:snrbanks}

In a matched-filter search for gravitational-wave signals, the signal
parameters $\boldsymbol\theta$ will not be known in advance.
The optimal detection statistic would be obtained by marginalizing~\cite{everitt2002the} the 
likelihood ratio $\Lambda({\bf d}|\boldsymbol{\theta})$ over all unknown parameters by 
integrating the likelihood ratio over these parameters\footnote{The integration measure to 
obtain an optimal statistic is given by the probability density of gravitational-wave 
signals over the unknown parameters~\cite{Searle:2008jv}: for example if the parameters 
$\boldsymbol{\theta}$ include $\iota$, the inclination of the binary orbit relative to the 
line of sight for a compact binary gravitational-wave source, the signal probability density 
over $\iota$ is uniform in $\cos \iota$~\cite{Schutz:2011tw}}.

Since the log likelihood ratio is a linear function of the signal model, its exponential -- 
the likelihood ratio itself -- will generally be sharply peaked about its maximum, thus the 
maximum value of $\Lambda({\bf d}|\boldsymbol{\theta})$ over unknown parameters 
$\boldsymbol{\theta}$ is expected to be a good approximation to the marginalized likelihood 
ratio (up to a possible constant rescaling).  This maximization procedure is equivalent
to minimizing the residuals seen in the detector, which can be seen as
follows: The log likelihood can be written as
\begin{equation}
   \log \Lambda({\bf d}|\boldsymbol{\theta})  =
-\frac12({\bf d} - {\bf h}(\boldsymbol{\theta}) \mid ({\bf d} - {\bf h}(\boldsymbol{\theta}))
+\frac12 ({\bf d} \mid {\bf d}).
\end{equation}
Now it is clear that the parameters $\hat{\boldsymbol\theta}$ that maximize
the log likelihood ratio are those that minimize the residuals
${\bf d} - {\bf h}(\boldsymbol{\theta})$ in terms 
of the noise weighted inner product.

The parameters ${\boldsymbol\theta}$ describing the strain observed in
a detector
include the signal amplitude $A$ observed in the detector
(which is inversely
proportional to the distance to a gravitational-wave source), the phase $\phi$
of the sinusoidally-varying signal observed in the detector, the arrival time
$t$ of the signal (usually defined by the moment when it reaches peak gravitational-wave amplitude at the detector), and
other parameters $\boldsymbol\mu$ describing the physical parameters of
the source such as the masses and spins of the components.  We write
\begin{equation}\label{e:hdecomp}
{\bf h}({\boldsymbol\theta}) =
A {\bf p}(t,{\boldsymbol\mu}) \cos\phi
+ A {\bf q}(t,{\boldsymbol\mu}) \sin\phi
\end{equation}
where
${\bf p}(t,{\boldsymbol\mu})$ and
${\bf q}(t,{\boldsymbol\mu})$ are in-phase (cosine) and quadrature-phase (sine)
waveforms, normalized so that
$({\bf p}\mid{\bf p})=({\bf q}\mid{\bf q})=1$, and which are orthogonal,
$({\bf p}\mid{\bf q})=0$.


Maximization over the amplitude and phase can be done algebraically
as follows:
Eq.~(\ref{e:lnlambda}) can be rewritten using Eq.~(\ref{e:hdecomp}) as
\begin{equation}
\log \Lambda({\bf d}|\boldsymbol{\theta})  =
A \rho(t,{\boldsymbol\mu}) \cos(\phi - \varphi) - \frac12 A^2
\end{equation}
where
\begin{equation}
\varphi \equiv \arctan\frac{
	({\bf d} \mid {\bf q}(t,{\boldsymbol\mu}))
}{
	({\bf d} \mid {\bf p}(t,{\boldsymbol\mu}))
}
\end{equation}
and
\begin{equation}
\rho(t,{\boldsymbol\mu}) \equiv \sqrt{
	({\bf d} \mid {\bf p}(t,{\boldsymbol\mu}))^2
	+ ({\bf d} \mid {\bf q}(t,{\boldsymbol\mu}))^2
}
\end{equation}
is the \emph{SNR time series}
for waveform templates with parameters $\boldsymbol\mu$.
The log-likelihood $\log\Lambda$ is maximal for amplitude
$\hat{A} = \rho$ and phase $\hat{\phi} = \varphi$ with
\begin{equation}
\max_{A,\phi} \log \Lambda({\boldsymbol\theta}) \equiv
\log \Lambda(t,\hat{A},\hat{\phi},{\boldsymbol\mu}) =
\frac12 \rho^2(t,{\boldsymbol\mu}) ~ .
\end{equation}
Peaks in this time series correspond to times at which a signal is most likely to be present.
Under the signal (noise) hypothesis, and in the presence of stationary and
Gaussian noise with a known power spectrum,
$\rho^2(t_{\text{peak}},{\boldsymbol\mu})$
follows a non-central (central) chi-squared distribution with two degrees of
freedom.

The SNR time series can be conveniently expressed in terms
of a complex time series as Eq.~(\ref{nwip}) of~\cite{Allen:2005fk}
\begin{equation}\label{complexsnr}
z(t,{\boldsymbol\mu}) = 4 \int_0^\infty \frac{\tilde{d}(f)\tilde{p}^\ast(f,{\boldsymbol\mu})}{S_n(f)}e^{2\pi ift} df
\end{equation}
as $\rho = |z|$ and the phase $\varphi = \arg(z)$; $\tilde{p}(f,{\boldsymbol\mu})$ is the Fourier transform of the in-phase waveform (see Eq.~\ref{e:hdecomp}).  Eq.~(\ref{complexsnr})
is the inverse Fourier transform of
\begin{equation}
\tilde{z}(f,{\boldsymbol\mu}) = 4 \frac{\tilde{d}(f)\tilde{p}^\ast(f,{\boldsymbol\mu})}{S_n(f)} \Theta(f)
\end{equation}
where $\Theta(f)$ is the Heaviside step function.  

Parameters such as the masses and spins of the binary components (represented
by $\boldsymbol\mu$) change the
morphology of the gravitational wave. 
In order to detect signals with a wide range of possible masses and spins, a bank consisting of large numbers of signal templates spanning the parameter space is produced, and each template in the bank is used as a matched filter.
The template bank used to initially find signals in the data is constructed
with a density in parameter space sufficiently high that the loss
in SNR between a true signal and the best-fit template is
less than 3\%.  
See~\cite{TheLIGOScientific:2016qqj,TheLIGOScientific:2016pea,LIGOScientific:2018mvr} for more details on the template banks used for the O1 and O2 searches.


An important property of the SNR should be noted: in
Eq.~(\ref{complexsnr}) it can be seen that the integrand is proportional to
$\tilde{d}(f)/S_{n}(f)$, so the data are not simply being whitened (which would
have been the case if the denominator were $S^{1/2}(f)$), but in fact
noisier parts of the frequency spectrum (including narrow lines) are \emph{suppressed} in the
matched filter.
Equivalently, the SNR integral can be seen as correlating a whitened data time series with a whitened template.
The SNR  therefore provides a natural way to
down-weight the frequency bands where the noise is large,
and effectively notches out the various lines.


\subsection{Rejection of noise artifacts and construction of candidate ranking statistic}
\label{ss:artifacts}

While the SNR  is the optimal detection statistic in the
case of stationary Gaussian noise, transient instrumental artifacts make it
a non-optimal statistic with real detector noise.
Although the matched filter naturally suppresses stationary noisy features in
the data, glitches can cause certain templates to produce high SNR
values~\cite{Christensen:2004kh,Christensen:2005gh,0264-9381-27-19-194010,TheLIGOScientific:2017lwt}. We address this in several different
ways:
\edef\keepparindent{\the\parindent}\relax
\begin{enumerate}
\item As explained in Section~\ref{ss:DQ-Terr-Noise}, we use witness sensors to identify times when the
environment or the instrument introduces frequent glitches and we veto
a subset of these times found to impact search performance from our analysis~\cite{TheLIGOScientific:2017lwt}. These sensors include those that monitor the physical environment about the gravitational-wave detector, as well as those that record signals from within the internal control systems of the interferometer.
\item We implement waveform consistency tests which characterize the deviation of
the data ${\bf d}$ from the model
${\bf n}+{\bf h}$~\cite{Allen:2004gu,Allen:2005fk,Usman:2015kfa,Messick:2016aqy}.
For signals from compact binary mergers, these tests are extremely powerful and
allow us to reject many glitches which have not been identified and vetoed, though
for short signals the discriminatory power of these tests is diminished~\cite{Allen:2004gu,Nitz:2017lco,Usman:2015kfa,LIGOScientific:2018mvr}.

\hspace*{\keepparindent}%
The exact implementation of these signal consistency tests vary among search
pipelines, but all are based on the following principle: if the gravitational-wave
model waveform is subtracted from the data to produce residuals
${\bf d}-{\bf h}$, the residuals should be consistent with Gaussian noise 
if the signal hypothesis is true.  These residuals are re-filtered
with the matched filter over different time or frequency intervals to determine
if non-noise-like features persist; evidence of such features suggest that the
model waveform ${\bf h}$ is not a good match to the non-Gaussian feature in the
data, and the detection ranking statistic is down-weighted accordingly.

\hspace*{\keepparindent}%
For example, the consistency test described in~\cite{Allen:2004gu,Allen:2005fk}
constructs a chi-squared test statistic by dividing the matched filter into
$n$ frequency bands as
\begin{equation}
  \chi^2 = \sum_{i=1}^{n}
  \frac{
     |({\bf d}-{\bf h} \mid {\bf p})_i - ({\bf d}-{\bf h} \mid {\bf p})/n|^2
     + |({\bf d}-{\bf h} \mid {\bf q})_i - ({\bf d}-{\bf h} \mid {\bf q})/n|^2
  }{1/n}
\end{equation}
where $({\bf a}\mid{\bf b})_i$ is the same as the inner product in
Eq.~(\ref{nwip}) but with the integrand restricted to the frequency interval
$f_{i-1}<f<f_{i}$ with $f_0=0$ and $f_n=\infty$.  Here the bands are
chosen so that $({\bf p}\mid{\bf p})_i=({\bf q}\mid{\bf q})_i=1/n$.
If the residual ${\bf d}-{\bf h}$ is Gaussian noise, $\chi^2$ is
chi-squared distributed with $\nu=2n-2$ degrees of freedom;
values of $\chi^2\gg\nu$ are indicative of residual non-Gaussian features
in the data after the model has been subtracted.  A re-weighted ranking
statistic proposed in~\cite{Usman:2015kfa}
\begin{equation}
   \hat{\rho} = \rho \times \left\{
   \begin{array}{ll}
      1 & \chi^2 \le \nu \\
      \left[\frac12 + \frac12(\chi^2/\nu)^3\right]^{-1/6} & \chi^2 > \nu
   \end{array}
   \right.
\end{equation}
down-weights the SNR for large values of $\chi^2$.  A similar time-domain based
signal consistency test is described in~\cite{Messick:2016aqy} and is
incorporated into a likelihood ranking statistic. 
\item For all detections published to date we have required that gravitational-wave signals be identified via matched
filtering in at least two independent detectors with consistent parameters. For
example, the arrival times of the gravitational waves at each detector must
differ by no more than the the maximum time-of-flight between the detectors, 
e.g.\ 10\,ms for the LIGO Hanford - LIGO Livingston pair, with an extra 5\,ms added in order to 
account for uncertainty in the inferred coalescence time at each detector. 
This 5\,ms addition to the coincidence window is also used when searching for simultaneous events for the LIGO Hanford - Virgo pair (27\,ms light travel time), 
and the LIGO Livingston - Virgo pair (26\,ms light travel time)~\cite{LIGOScientific:2018mvr}.
However, having now established the existence and frequency of gravitational-wave signals,
it may now also be possible to make detections when only one detector is operating, 
and thus this time coincidence test is not available~\cite{Callister:2017urp}. 
\end{enumerate}

The matched-filter based searches employed by the LVC construct
ranking statistics from the SNR and the waveform consistency
test statistics~\cite{LIGOScientific:2018mvr}.
In addition
an astrophysical signal received in several detectors will have a common set of
parameters $\boldsymbol\mu$ (within limits imposed by limited SNR) in all detectors, and, furthermore, the amplitude,
phase, and time-of-arrival of the signals observed in each detector will be
determined by the direction of propagation of the wave (i.e., from where on the
sky the signal originates) and the polarization state of the signal.
Since gravitational waves have two polarizations (in general relativity),
referred to as the plus-polarization ${\bf h}_+$ and the cross-polarization
${\bf h}_\times$, the strain on detector $I$ is determined by the detector's
antenna response patterns $F_{+,I}$ and $F_{\times,I}$ by
\begin{equation}
{\bf h}_I({\boldsymbol\theta}) =
F_{+,I}(\alpha,\delta,\psi,t) {\bf h}_+(t-\tau_I,D,\iota,{\boldsymbol\mu}) + 
F_{\times,I}(\alpha,\delta,\psi,t) {\bf h}_\times(t-\tau_I,D,\iota,{\boldsymbol\mu})
\end{equation}
where $\alpha$ and $\delta$ are the right ascension and declination of the
source of the gravitational waves, $D$ is the distance to the source of
the waves, $\iota$ is the inclination of the orbital plane of the binary system (which, for circular
orbits and leading order quadrupole emission, determines the ellipticity angle), and 
$\tau_I=\tau_I(\alpha,\delta,t)$ is the travel time of the signal
from the geocenter to the detector. Although a fully-coherent search for
gravitational waves across a network of detectors is possible, we opt instead
to perform searches independently in each detector and then demand that
triggers seen in different detectors have consistent times of arrival and the
same parameters ${\boldsymbol\mu}$ since this provides a powerful glitch
rejection consistency test as described above.  However, further signal
consistency requirements are also possible.  For the leading-order quadrupole
emission from a circular binary, the ratios of the amplitudes seen in two
detectors is
\begin{equation}
\frac{A_I}{A_J} = \sqrt{
\frac{\displaystyle
F_{+,I}^2\left(\frac{1+\cos^2\iota}{2}\right)^2+F_{\times,I}^2\cos^2\iota
}{\displaystyle
F_{+,J}^2\left(\frac{1+\cos^2\iota}{2}\right)^2+F_{\times,J}^2\cos^2\iota
}}
\end{equation}
while the difference in arrival time is $t_I - t_J = \tau_I - \tau_J$ and the
difference in phase is
\begin{equation}
\phi_I - \phi_J =
\arctan\left(\frac{F_{\times,J}}{F_{+,J}}\frac{2\cos\iota}{1+\cos^2\iota}\right)
-
\arctan\left(\frac{F_{\times,I}}{F_{+,I}}\frac{2\cos\iota}{1+\cos^2\iota}\right)
.
\end{equation}
It is therefore
possible to include an amplitude-phase-time consistency measure in
likelihood-based ranking statistics~\cite{Usman:2015kfa,2015arXiv150404632C,TheLIGOScientific:2017lwt}; this was done for the most recent searches for gravitational-wave signals in the LIGO-Virgo O1 and O2 data~\cite{LIGOScientific:2018mvr}. 


\subsection{Background estimation and detection confidence}
\label{ss:background}

After the steps described above which mitigate the effects of noise transients,
the probability of remaining transients occurring simultaneously (within a time window that takes into account the maximal travel time of a signal, for example 10\,ms for the two LIGO detectors)
in two detectors and producing a large joint ranking statistic value
becomes extremely small.  Different searches adopt different approaches for
measuring this probability as a function of the ranking statistic~\cite{TheLIGOScientific:2017lwt}.
The basic method is to examine the statistical properties of the non-simultaneous transients observed in each detector and to artificially treat them as if they did occur simultaneously.  

The statistical significance of any candidate event observed in two or more detectors is quantified by its \emph{false alarm rate}, which is the expected rate of events per time due to noise which would be assigned an equal or larger ranking statistic than the candidate.
One approach to estimating false alarm rates is to shift one detector’s data stream in time (by a time interval larger than the maximum time-of-flight between detectors) and repeat the search.  The resulting ``time-shifted'' coincidences are then treated as a background noise sample.
This is done numerous times with different time shifts in order to obtain a probability distribution for
the joint detector ranking statistics. 
Each coincident trigger is assigned a false alarm rate given
by the number of background triggers with an equal or larger
ranking statistic,  divided  by  the  total  time  searched  for  time-shifted coincidences.
For example, in~\cite{Abbott:2016blz} it is found that
the frequency of transients producing more significant events than GW150914
is less than once every 200\,000 years in both of the matched-filter
searches employed by the LVC.

Another similar approach is to accumulate single-detector triggers not having
simultaneous (within the time-of-flight of gravitational waves) triggers in another detector and therefore likely not
associated with gravitational-wave signals.
The distribution of the ranking statistic under the background (noise only) hypothesis is then estimated by randomly drawing
single detector triggers from the inferred
single detector distributions and artificially treating them as if they were
simultaneous when constructing the ranking statistic. 
The significance
of an observed ranking statistic value can then be evaluated using this
background distribution~\cite{Messick:2016aqy,2015arXiv150404632C}.
These two independent methods of determining the significance of observed
gravitational-wave candidates have both yielded high significance for the
gravitational waves that have been identified.

Terrestrial noise sources that are potentially correlated between detectors are not taken into account by these background estimation methods. 
Thus, as discussed in Section~\ref{ss:DQ-Terr-Noise} above, a detailed examination of physical and environmental sensors which monitor such 
noise sources and an assessment of their coupling to the measured gravitational-wave strain channel is carried out in order to check the 
validity of a detection candidate.

The LVC also performs searches of LIGO and Virgo data without specific
waveform models~\cite{Klimenko:2015ypf,Klimenko:2008fu}.  These searches
first identify periods of excess power in each detector's data stream and
then builds a detection statistic based on the cross-correlation between the
data streams.  The significance of a particular detection statistic value
is again assessed using time shift analyses. 
 In~\cite{Abbott:2016blz}
it was shown that the frequency of noise transients producing more significant
events than GW150914 in such a generic transient search is once every
8\,400 years.  

The fact that multiple searches, employing different methods, all found
GW150914 to be a highly significant candidate bolsters our confidence that
this event is not the product of coincident transient noise.
Furthermore, the signal is well matched by the waveform predicted by general
relativity for the coalescence of a binary black hole system.
Various tests performed using the first ten binary black hole mergers
detected by the LVC with high confidence have shown no significant deviation from general relativity models~\cite{LIGOScientific:2019fpa}.

\subsection{Measurement of search sensitivity}

A final component of a search pipeline is the determination of the sensitivity
of the search to astrophysical populations of signals.  The purpose of
doing so is two-fold:  First, it provides a metric by which a search can
be tuned to optimize its detection efficiency for particular classes of
signals; second, it provides a means for interpreting the rate of signals
detected by the pipeline to the rate at which signals are generated by
the population.

The sensitivity of the search pipeline is normally measured via a 
Monte Carlo procedure (see, e.g., \cite{Tiwari:2017ndi}) in which
simulated signals drawn from a hypothetical source population (e.g.,
some distribution of binary component masses and spins, orientation angles,
arrival times, and distance) are added to real detector
noise, and the search is rerun to determine the fraction of signals
from this population that are detected by the search pipeline.  A simulated
signal is considered to be detected if the search pipeline produces a
trigger above some chosen ranking statistic threshold.  The result is
represented as a time- and population-averaged spacetime sensitivity
$\langle VT\rangle$ for a fixed ranking statistic threshold which corresponds
to a fixed false alarm rate threshold (e.g., a threshold could be chosen
to be one false alarm per century of observation).  Alternatively,
threshold-independent
methods of astrophysical rate estimates can also be employed
\cite{Farr:2013yna,Abbott:2016nhf,Abbott:2016drs}.

%% file: pe.tex
\section{Inferring waveform and physical parameters}
\label{s:pe}

Once a candidate gravitational-wave signal is identified, and its significance is established,
the next goal is to use the data to infer the physical parameters of the system that created the gravitational waves~\cite{PhysRevD.58.082001,PhysRevD.64.022001,Veitch:2009hd,Veitch:2014wba,Rover:2006ni,vanderSluys:2008qx,vanderSluys:2007st,TheLIGOScientific:2016wfe}.  
The detection of gravitational waves
as well as the inference of the physical parameters relies on knowledge of the generic 
shape of the signal one is looking for as well as the distribution of the noise. 
Moreover, gravitational-wave signals are weak, therefore uncertainties in these parameters may be large 
and \emph{a priori} assumptions about the typical amplitudes and phase evolution of such signals do have a significant impact on the 
reconstructed waveform.
For these reasons, inference of the physical parameters of the system, such as masses, 
spins of the merging objects, is done within the framework of Bayesian parameter estimation. 
The central elements that need defining are a \emph{model} $M$ for the gravitational-wave signal that 
allows for the prediction of the form of the signal from the values of the physical parameters of the system, and the 
so-called \emph{background} or \emph{prior} information $I$. 

Given a model $M$ that depends on a set of parameters $\boldsymbol{\theta}$, background information $I$, 
and a set of observations (data) ${\bf d}$, inference is done via application of Bayes' theorem:
\begin{equation}\label{eq:bayes}
p(\boldsymbol{\theta} | {\bf d}, M, I) =p(\boldsymbol{\theta} | M, I)\frac{p({\bf d} | \boldsymbol{\theta}, M, I)}{p({\bf d} |M,  I)}\,.
\end{equation}
The left hand side is referred to as the \emph{posterior probability density function}, or simply the posterior for $\boldsymbol{\theta}$, 
while the three terms on the right hand side are the \emph{prior probability density function} $p(\boldsymbol{\theta} |M, I)$, 
the \emph{likelihood function} $p({\bf d}|\boldsymbol{\theta},M\, I)$, given by equation (\ref{like2}) and the \emph{evidence}, 
\begin{equation}\label{eq:evidence}
p({\bf d} |M,  I) = \int d \boldsymbol{\theta}\, p(\boldsymbol{\theta}|M,  I) p({\bf d} |\boldsymbol \theta, M, I)\,.
\end{equation}
Within the Bayesian parameter estimation framework, the inference is reduced to the calculation of the posterior 
for $\boldsymbol{\boldsymbol{\theta}}$ given the model $M$ and the analysis assumptions $I$ which uniquely determine 
the prior distribution and the likelihood function.

\subsection{Waveform models} \label{waveforms}

Let us focus now on the choice of signal model $M$. The signal model 
$M$ determines the functional form of $h(t;\boldsymbol{\theta})$ which is key to calculating the 
likelihood function. For definiteness, we will concentrate on 
parametric forms of $h(t;\boldsymbol{\theta})$, as obtained by solving Einstein's equations. 
For a discussion of non-parametric signal models see~\cite{Cornish:2014kda}. 
Exact analytic solutions of Einstein's equations are notoriously difficult to obtain; 
therefore data-analysis-ready models are either based on perturbative solutions, 
e.g.\ the Taylor family of waveforms~\cite{Buonanno:2009zt} or the effective-one-body 
waveforms~\cite{Buonanno:1998gg,Damour:2001tu,Taracchini:2013rva,Bohe:2016gbl,Nagar:2018zoe}, 
or on hybrid/phenomenological approaches such as the Phenom family of waveforms
~\cite{Ajith:2009bn,Hannam:2013oca,Husa:2015iqa,Khan:2015jqa}. 
We will not discuss further the details of the waveform models here, but we will restrict ourselves to
the two main types of waveforms employed in the original analysis of GW150914. 
For GW150914, the two models used in~\cite{TheLIGOScientific:2016wfe} were SEOBNRv2 and 
IMRPhenomPv2~\cite{TheLIGOScientific:2016wfe,lalsimulation}. Both waveform models are full inspiral-merger-ringdown models 
that succeed in reproducing numerical waveforms, especially in the region of approximately equal mass
and moderate spins magnitudes. The main difference between the two models lays in the treatment of the spin dynamics. 
SEOBNRv2 models the dynamics of the component of the spin vectors along the direction of the orbital angular momentum
while IMRPhenomPv2 includes also an effective treatment of the dynamics of the in-plane components of the spins, 
and thus includes an approximate precessing dynamics~\footnote{At the time of the discovery of GW150914 another precessing waveform model
was available, SEOBNRv3, which also includes in-plane spin components~\cite{Pan:2013rra,Babak:2016tgq}. The original analysis, however,
did not include results from this model, which were reported in~\cite{Abbott:2016izl}.}. 
Because GW150914 was nearly a face-off system (orbital angular momentum vector pointing away from the Earth), the LIGO instruments were not
sensitive to the in-plane spin components, hence the two waveform models 
are essentially equivalent.
In~\cite{Abbott:2016wiq}, the LVC has empirically shown that the inferred properties of GW150914
depend relatively weakly on a change in the waveform model.
This finding was confirmed by the analysis presented in~\cite{Nagar:2018zoe} 
using an independent effective-one-body implementation and by~\cite{Abbott:2016apu}, in which numerical relativity solutions were directly compared with 
GW150914 data.

\subsection{Prior distributions}
The final functions necessary for the application of Bayes' theorem, Eq.~(\ref{eq:bayes}), are
the prior probability distributions for the parameters of interest. These are all the parameters necessary 
to completely characterise the gravitational-wave signal emitted during a coalescence event. For quasi-circular orbits,
these are:
\begin{itemize}
\item the component masses $m_1$ and $m_2$;
\item the spin vectors $\vec{S}_1$ and $\vec{S}_2$;
\item the polarisation angle $\psi$ and the angle $\theta_{jn}$ between the total angular momentum $\vec{J}$ and the 
propagation direction of the gravitational wave $\hat{n}$;
\item the source luminosity distance $D_L$;
\item the source right ascension $\alpha$ and declination $\delta$;
\item a reference phase $\varphi_0$ and a reference time, typically the gravitational-wave strain peak time, $t_0$. 
\end{itemize}
The functional form of the prior distribution must be specified for all parameters.  
In some cases the prior distribution is determined via invariance (symmetry) properties of the parameter space~\cite{royden2017real}; for instance, the prior for the source position 
$D_L, \alpha, \delta$ in the Universe is
chosen from the requirement that the number density of sources is uniform in the cosmological co-moving volume in accordance with
a Friedmann-Lema\^itre-Robertson-Walker cosmological model. Thus, the probability $p(D_L, \alpha, \delta|M\, I) \propto dV $
and for redshift $z \ll 1$ reduces to $p(D_L, \alpha, \delta|M\, I) \propto D_L^2 \vert \cos(\delta)\vert$. 
In cases where invariance arguments do not apply, we choose simple forms of prior distribution so that the resulting posteriors are easily interpretable.
Similar arguments determine the prior for the spin vectors $\vec{S}_1, \vec{S}_2$ and orientation angles $\psi,\theta_{jn}$ 
to be uniform over the azimuthal angles ranging between $0$ and $2\pi$ as well as uniform in the cosine of the polar angles ranging between $-1$ and $1$.
Regarding the spin vector magnitudes, several possible priors are possible, e.g.\ $p(|\vec{S}_i||M,I) \propto |\vec{S}_i|^2$ or   
$p(|\vec{S}_i||M,I) \propto 1$. The main analysis of the events in the GWTC-1 catalog employed a uniform distribution over the norm of the spin vectors.
For the component masses $m_1$ and $m_2$, the chosen prior distribution is uniform, thus $p(m_1, m_2 |M\, I) \propto 1$, but 
limited from below so that $m_1,m_2 > 1M_{\odot}$. 

\subsection{Calibration uncertainties}
In addition to uncertainties induced by detector noise, the accuracy and precision of our source parameter estimates are also 
affected by uncertainties in the amplitude and phase response of the detectors.
For this reason, this source of uncertainty on the data is modelled and included in the analysis. The calibration uncertainty model
employed is based on empirical estimates of the error magnitudes in both amplitude and phase in specified frequency bands~\cite{Abbott:2016jsd,Cahillane:2017,Acernese:2018bfl}. In particular, 
the model assumes the value of the errors to be distributed as a Gaussian distribution with zero mean and a variance 
given by the empirically determined error magnitudes. Calibration uncertainty curves
are then constructed using third order spline interpolation over the data frequency space. Typically, this introduces a total 
of $O(10)$ additional parameters per detector (half for the phase uncertainty and the other half for the amplitude), which 
are sampled in concert with the physical parameters of the system. Technical details for the LVC calibration model can be found in~\cite{calibrationerrors}.

\subsection{Numerical methods}

The total number of parameters to be inferred is thus 15 for quasi-circular orbits and generic spin vectors, and 11 for models where spins are 
forced to be aligned with the orbital angular momentum. 
To the set of physical parameters, we must add the 10 parameters per detector necessary to specify the calibration uncertainty model. Hence, 
for a typical three-detector analysis, we sample a grand total of 45 parameters for a quasi-circular system with generic spin orientations.
Parameter spaces of such high dimensionality cannot be efficiently explored with grid-based methods. Therefore, 
over many years members of the LVC developed the \texttt{LALInference} stochastic sampler library~\cite{Veitch:2014wba} 
which implements two algorithms, a parallel tempering Markov chain Monte Carlo~\cite{Rover:2006bb} and a nested sampling~\cite{Veitch:2009hd}.
The parallel tempering Markov chain Monte Carlo is designed to generate samples from the multidimensional posterior distribution (\ref{eq:bayes}), while the nested sampling
instead is designed to calculate the evidence, Eq.~(\ref{eq:evidence}) and generates samples from the posterior distribution as 
a by-product. More details are given in~\cite{Veitch:2014wba} and references therein. 
Other parameter estimation pipelines are routinely used by the LVC, such as rapidPE~\cite{Lange:2018pyp} and BILBY~\cite{Ashton:2018jfp}, but for the
rest of the discussion we will focus on \texttt{LALinference}. However, the same considerations will apply to other Bayesian analysis methods.

\subsection{Posterior distributions}

\begin{figure}[h]
\centering \includegraphics[scale=0.5]{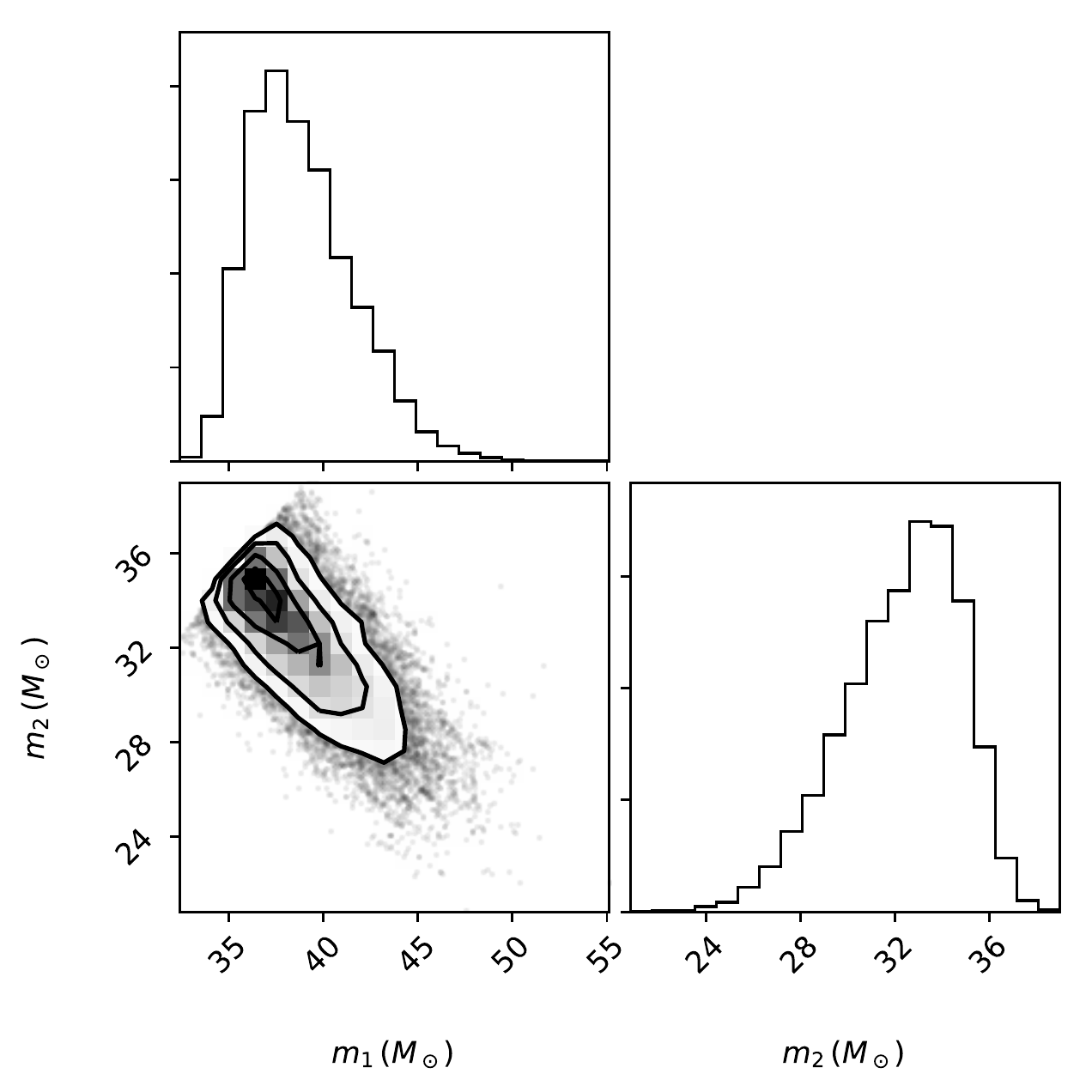} 
\caption{\label{fig:m1-m2} One- and two-dimensional posterior distribution for the detector-frame masses for the
GW150914 event obtained using the IMRPhenomPv2 waveform model. The three panels show (i) the one-dimensional marginal posterior for $m_1$ (top panel); (ii) the joint two-dimensional $m_1 - m_2$ posterior (bottom left); (iii) the one-dimensional marginal posterior for $m_2$ (bottom right). Posterior samples taken from~\cite{PEsamps}.}
\end{figure}

The end products of the \texttt{LALInference} analyses are posterior samples for all parameters that characterise the gravitational-wave waveform. Of particular interest are posteriors for the \emph{intrinsic parameters}, masses and spin vectors, which help ascertain the nature of the coalescing objects.
For GW150914, the detector-frame masses (i.e., redshifted due to cosmological expansion) measured using the IMRPhenomPv2 waveform model were $38.5_{-3.6}^{+5.6} M_\odot$ and $32.2_{-4.8}^{+3.6} M_\odot$; see Table I in~\cite{TheLIGOScientific:2016wfe}. The quoted numbers are an extremely concise way of summarising the full posterior distribution. The output of \texttt{LALInference} analyses are samples from the \emph{full} posterior distribution. In particular, a number such as  $38.5_{-3.6}^{+5.6} M_\odot$ comes from the marginalisation over 14 of the 15 physical source parameters of the full posterior, as well as the calibration parameters, to obtain a one-dimensional posterior from which the 90\% credible region is then calculated. Naturally, correlations between different parameters are invisible in a one-dimensional representation. For a clearer picture, multidimensional posterior distributions help to display the information extracted from the analysis. Figure~\ref{fig:m1-m2} shows the joint two-dimensional posterior distribution for the component masses $m_1$ and $m_2$ as an example.
In particular, the bottom left panel shows the non-negligible correlation between the component masses.

The full details of the multi-dimensional posterior distribution can be visualised in a compact way by computing the posterior distribution over the waveform itself in the time domain. This is done simply by computing the predicted waveform over each of the posterior samples. Let $\boldsymbol{\theta}_i$ be the $i$-th posterior sample, the corresponding waveform will be $h(t;\boldsymbol{\theta}_i)$. The waveform samples are the set $\{h(t;\boldsymbol{\theta}_i)\}_{i=1,\dots,N}\equiv \{h_i\}$. Each of the waveform samples $h_i$ can be whitened, see Section \ref{sec:noise}, and then used to compute credible intervals at every time $t_j$ at which the original data were sampled. The result of this procedure is summarised in the presentation of Figure 6 of~\cite{TheLIGOScientific:2016wfe}. 
Figure 1 in~\cite{Abbott:2016blz} is representing a different procedure; the second row in this figure shows a comparison between the reconstructed 90\% credible region obtained by the procedure described above, and a numerical relativity solution that, while not corresponding to any of the computed posterior samples, is consistent with the reconstructed 90\% credible region. 


\subsection{Validation of source parameter estimates}

The results from Bayesian inference are only as good as the models used in the analysis. If the waveforms used in the signal model or the underlying assumptions of the noise model are inaccurate, the results will suffer from systematic bias. 
A multitude of tests are used to check for possible mis-modeling error and to quantify the impact on the analyses. As discussed earlier in Section~\ref{waveforms} the waveform models are compared to highly accurate numerical relativity simulations, and multiple waveform approximants are used in the analyses and cross-compared. The difference between the results found using different waveform models provides an estimate of the systematic error due to the signal model. The noise model can also be checked.
Other checks include
adding simulated signals with similar parameters to the astrophysical events into nearby stretches of data and checking that the parameters are properly recovered by the parameter estimation algorithms.

\begin{figure}[h]
\centering \includegraphics[scale=0.33]{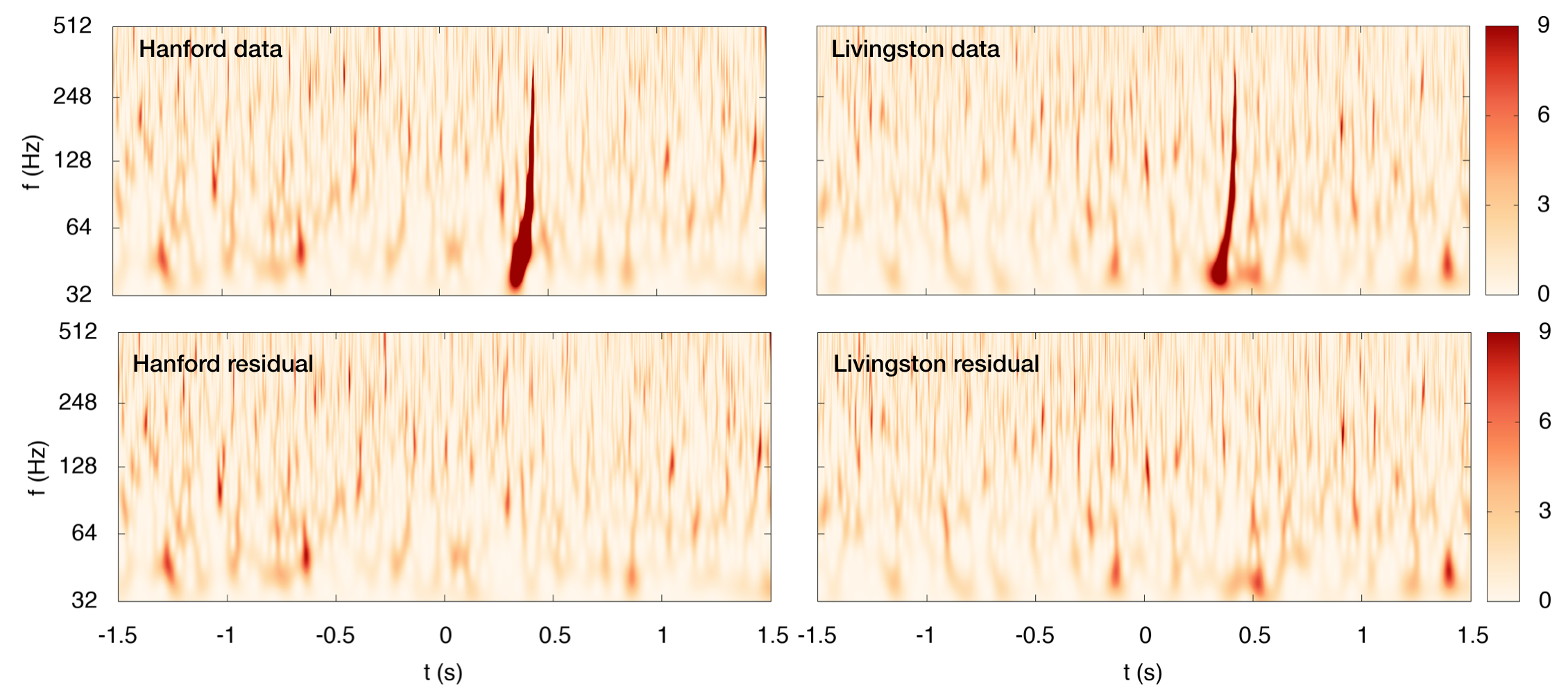} 
\caption{\label{fig:qscan} Scalograms (or Q-scans) of the whitened data and residuals in the LIGO-Hanford and LIGO-Livingston detectors in 3 seconds of data surrounding the GW150914 event. 
The residuals are free of glitches or correlated power. The color scale, as displayed by the bar on the right, corresponds to the whitened power.}
\end{figure}

Over long stretches of LIGO-Virgo data, the noise is known to be non-stationary and non-Gaussian. The overall noise levels fluctuate, and there are frequent low-SNR glitches, and less frequently high-SNR glitches, see Section~\ref{sec:timefreq}. On the other hand, the gravitational-wave signals spend very little time in the LIGO-Virgo sensitive band---seconds or less for black hole binaries and minutes for neutron star binaries, and over these shorter stretches of time the noise is generally (but not always) well approximated as stationary and Gaussian. When a significant trigger is found by the search pipelines, the first thing the analysts look at are multi-resolution time-frequency scalograms of the data surrounding the trigger (known as Q-scans). 
Q-scans are qualitative checks which require visual inspection~\cite{chatterji:2005,gravityspy}.
These scans reveal whether there are any loud glitches in the data, as was the case with the binary neutron star GW170817~\cite{TheLIGOScientific:2017qsa}. Once the parameter estimation analyses have been run, Q-scans of the residuals are closely examined to see if any any unmodelled noise features might have affected the analyses. Figure~\ref{fig:qscan} shows Q-scans of the whitened data and residuals surrounding GPS time 1126259462. The scans of the data reveal the signal from GW150914, while the residuals after subtracting the maximum likelihood waveform from the parameter estimation studies~\cite{TheLIGOScientific:2016wfe} show no visible evidence of glitches or correlated signal power.

In addition to these qualitative checks, more rigorous quantitative checks can be applied. One test that is routinely applied is to reanalyze the residuals using the wavelet-based BayesWave algorithm~\cite{Cornish:2014kda} which is able to identify any glitches and remaining coherent power. Coherent power in the residuals could be evidence of departures from general relativity, or evidence of shortcomings in the template models or the noise model used for parameter estimation. No significant coherent power was found in the residuals for any of the detected events. In the case of GW150914 the lack of a coherent residual was used to place interesting bounds on possible departures from general relativity~\cite{TheLIGOScientific:2016src}. In the case of the binary neutron star merger GW170817~\cite{TheLIGOScientific:2017qsa}, a loud incoherent glitch was seen to overlap the signal in the Livingston detector. The glitch was reconstructed and removed using the BayesWave algorithm. The glitch removal procedure has been shown to be safe in a study that injected simulated neutron star merger signals into data with similar loud glitches, followed by removing the glitches with BayesWave and accurately recovering the true signal parameters with the LVC parameter estimation algorithms~\cite{Pankow:2018qpo}.

\begin{figure}[h]
\centering \includegraphics[scale=0.8]{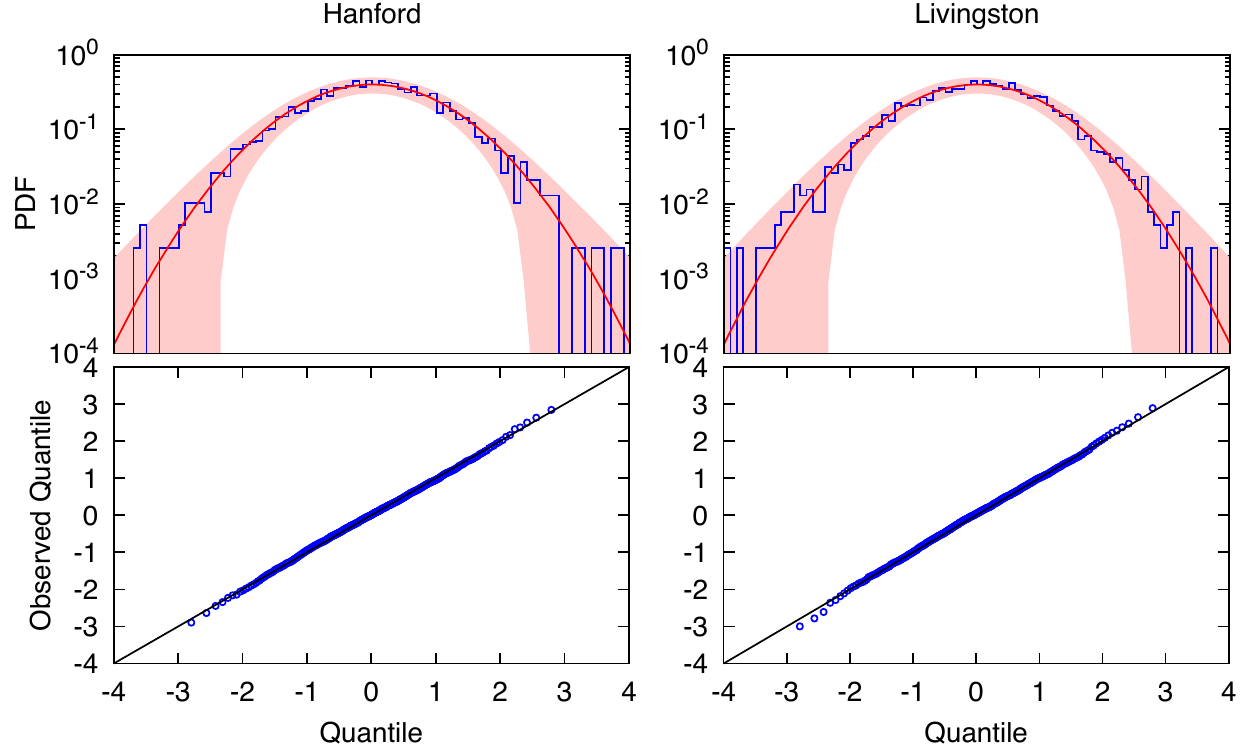} 
\caption{\label{fig:pe} Histograms and quantile-quantile plots of the whitened Fourier amplitudes of the residuals in the LIGO-Hanford and LIGO-Livingston detectors for 4 seconds of data surrounding GW150914. The shaded band in the upper panels indicates the expected 3-sigma variance from having a finite number of samples.The residuals show no evidence of non-Gaussianity.}
\end{figure}

Figure~\ref{fig:pe} shows histograms of the whitened Fourier amplitudes of the residuals in the LIGO-Hanford and LIGO-Livingston detectors following the removal of the maximum likelihood waveform for GW150914. The residuals are taken from the parameter estimation analysis published at the time of the discovery~\cite{TheLIGOScientific:2016wfe}. These residuals were used to test for residual coherent power, which can be framed as a test of general relativity if we assume the template to be a sufficiently accurate solution of the theory. Applying the Anderson-Darling test of normality to the residuals yields p-values of 0.15 for LIGO-Hanford and 0.11 for LIGO-Livingston, indicating that the residuals are consistent with the Gaussian noise model used to define the likelihood.

Even if the residuals had failed the formal tests of stationarity and Gaussianity discussed here, it would not necessarily imply that the parameter estimation would be strongly biased. When the noise deviates from the model the analysis will suffer systematic bias. But for this bias to be significant it has to be large compared to the statistical spread in the posterior distributions. Extensive studies using simulated signals added to real LIGO-Virgo data have shown that systematic errors due to deviations from the noise models are generally negligible compared to the statistical uncertainties~\cite{Raymond:2009cv,vanderSluys:2009bf,vanderSluys:2007st,Berry:2014jja, Abbott:2016wiq, Pankow:2018qpo}. One exception is when the simulated signals cover or overlap the times of glitches, in which case the biases can be large~\cite{Powell:2018csz}. When glitches are present, tools such as BayesWave need to be used to model and remove the glitches, ideally in concert with the parameter estimation.

\subsection{Parameter degeneracies and credible intervals}

\begin{figure}[h]
\centering \includegraphics[scale=0.5]{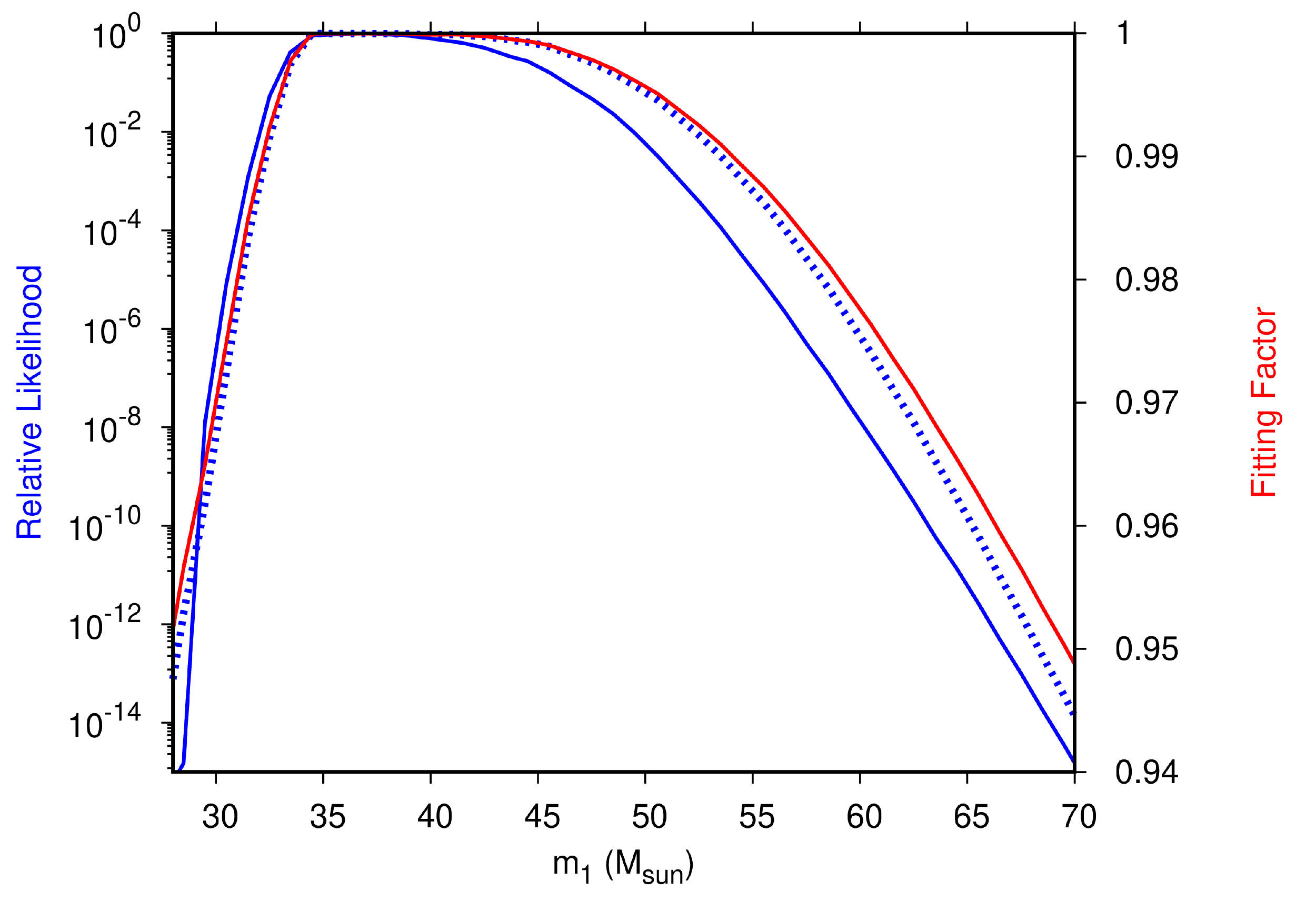}  
\caption{\label{fig:postlike}  The likelihood (in blue) and the fitting factor (in red) as a function of the detector-frame primary mass $m_1$ for GW150914. The dashed blue line shows the estimate of the likelihood in terms of the fitting factor from equation (\ref{logLFF}).
The likelihood is scaled relative to the maximum likelihood value. The maximization was performed over the secondary mass, spins and extrinsic parameters.}
\end{figure}

Gravitational-wave templates exhibit a variety of parameter degeneracies whereby templates with different parameters can have very similar amplitude and phase evolution, and yield very similar likelihoods. One example of such a degeneracy is evident in the posterior distribution for the component masses of GW150914 shown in Figure~\ref{fig:m1-m2}. The degree of similarity between templates with parameters $\boldsymbol{\lambda}, \boldsymbol{\theta}$ is measured by the {\em match}
\begin{equation}
{\rm M}(\boldsymbol{\lambda}, \boldsymbol{\theta})  =  \frac{ ({\bf h}(\boldsymbol{\lambda}) | {\bf h}(\boldsymbol{\theta}))}{\sqrt{({\bf h}(\boldsymbol{\lambda}) | {\bf h}(\boldsymbol{\lambda}) ) ({\bf h}(\boldsymbol{\theta}) | {\bf h}(\boldsymbol{\theta}))} }\, .
\end{equation}
If the true signal is described by ${\bf h}(\boldsymbol{\lambda})$, then the expectation value of the log likelihood for template ${\bf h}(\boldsymbol{\theta})$, maximized over amplitude is
\begin{equation}
{\rm E}[\ln\Lambda (\boldsymbol{\lambda}|\boldsymbol{\theta}) ] =  \frac{1}{2} {\rm M}^2(\boldsymbol{\lambda}, \boldsymbol{\theta})\,  {\rm SNR}^2 \, ,
\end{equation}
where $\rm SNR$ is the optimal signal-to-noise ratio~\cite{DelPozzo:2014cla}.
We see that signals that have similar morphology, as measured by the match, yield similar likelihoods. Now suppose we hold one parameter, $\theta^k$ fixed, then maximize the likelihood with respect to all the other parameters. Up to an overall constant, we have~\cite{Cornish:2011ys, DelPozzo:2014cla}
\begin{equation}\label{logLFF}
\ln\Lambda ({\bf d}| \bar\theta^k) \equiv {\rm max}_{j\neq k}\ln \Lambda({\bf d}| \theta^j) \simeq \frac{{\rm SNR}^2}{2}\,  {\rm FF}^2(\bar\theta^k)\, ,
\end{equation}
where ${\rm FF}(\bar\theta^k)$ is the {\em fitting factor}, or maximized match, between waveforms with $\theta^k=\bar\theta^k$ and the maximum likelihood waveform. 
Figure~\ref{fig:postlike} compares the maximum likelihood as a function of the primary detector-frame mass for GW150914 to the fitting factor. The fitting factor as a function of $m_1$ was computed by maximizing the match between the overall maximum likelihood waveform and waveforms with fixed $m_1$.
Note that the posterior distribution for this event had a 90\% credible interval of $m_1 = 38.5_{-3.6}^{+5.6}  M_\odot$, but templates with primary masses outside this interval continue to yield large fitting factors because other parameters can be adjusted to partly compensate the effects of the change in primary mass on the waveform. For example, we find a fitting factor between the maximum likelihood template for GW150914 and a template with a primary mass of $m_1=70 \, M_\odot$ of ${\rm FF}= 0.95$.

The possibility of achieving high matches, or correlations, between templates with large primary masses and the maximum likelihood template have been cited as evidence that the LIGO-Virgo parameter estimation analysis for GW150914 and other systems may be flawed~\cite{Creswell:2018tsr}. It was further hypothesized that the credible intervals were underestimated due to the instrument noise not conforming to the likelihood model~\cite{Creswell:2018tsr}. However, our confidence in the reconstructed credible regions comes from extensive simulations designed to compare the cumulative distributions of simulated populations against the cumulative distribution of reconstructed credible regions. The agreement between the two, see for instance Figure 10 in~\cite{Veitch:2014wba}, demonstrates that our algorithms are properly computing the credible intervals. 
Moreover, as we have shown, the noise properties for GW150914 {\em are} compatible with the likelihood model used in the parameter estimation studies, further reinforcing our 
confidence in the method used to compute the credible intervals. 

There is no contradiction in having templates with parameters outside the quoted credible regions producing large fitting factors with the best fit model, since even small template mismatches come with a large penalty for high signal-to-noise ratio systems such as GW150914. For example, the difference in log likelihood between the signal with $m_1=70 \, M_\odot$ and the global maximum is $\Delta \ln \Lambda = -32$, which is what we expect to see for a ${\rm SNR} \simeq 25$ signal and a template with a fitting factor of ${\rm FF}= 0.95$. But the relative likelihood for the higher mass solution is $e^{\Delta \ln \Lambda}=10^{-13.9}$, thus while templates with large primary masses can produce relatively good matches to the data, the probability that the primary mass is this high is vanishingly small.

%% file: residuals.tex
\section{Residuals analysis of LIGO data around GW150914}
\label{sec:res}

\setcounter{footnote}{0}

The notion of a residual -- the data minus the model -- plays an important role in gravitational-wave data analysis.
If the signal model matches well the true signal, then the residual should be consistent with a draw from the noise model 
$p({\bf n})$, the probability distribution for the noise. After known sources of correlation with independent witnesses are removed, we expect the instrument noise in the widely separated LIGO-Virgo detectors to be fully independent, and therefore the residuals in each detector to be uncorrelated.
In contrast, gravitational-wave signals will excite a coherent response across the network of detectors, and this difference in correlation properties is one of the ways we are able to separate signals from noise. 

As noted in Section~\ref{s:further-reading}, it is possible to have correlated transient noise due to lightning~\cite{0264-9381-34-7-074002}, but monitoring with magnetometers is presently adequate to rule that out as the cause of events like GW150914~\cite{TheLIGOScientific:2016zmo}.  Low-level correlated magnetic noise is more of a concern for the search for a stochastic gravitational-wave background~\cite{Abbott:2006zx,PhysRevD.87.123009,PhysRevD.90.023013,Coughlin:2018str}.  Seismic noise is similarly monitored.  Since the LIGO detectors share the same design and similar equipment, the frequencies associated with synchronized clocks (GPS), electrical power (60 Hz), and instrument resonances are monitored and suppressed in stochastic background and continuous-wave gravitational wave searches~\cite{Covas:2018}.

In this section we will use the data surrounding GW150914 to illustrate the discussion, but the same considerations apply in general, and analyses of the residuals have been reported for all significant events~\cite{LIGOScientific:2019fpa}.

\subsection{Signal and template comparisons} \label{ss:comparisons}

\begin{figure}[h]
\centering \includegraphics[scale=0.3]{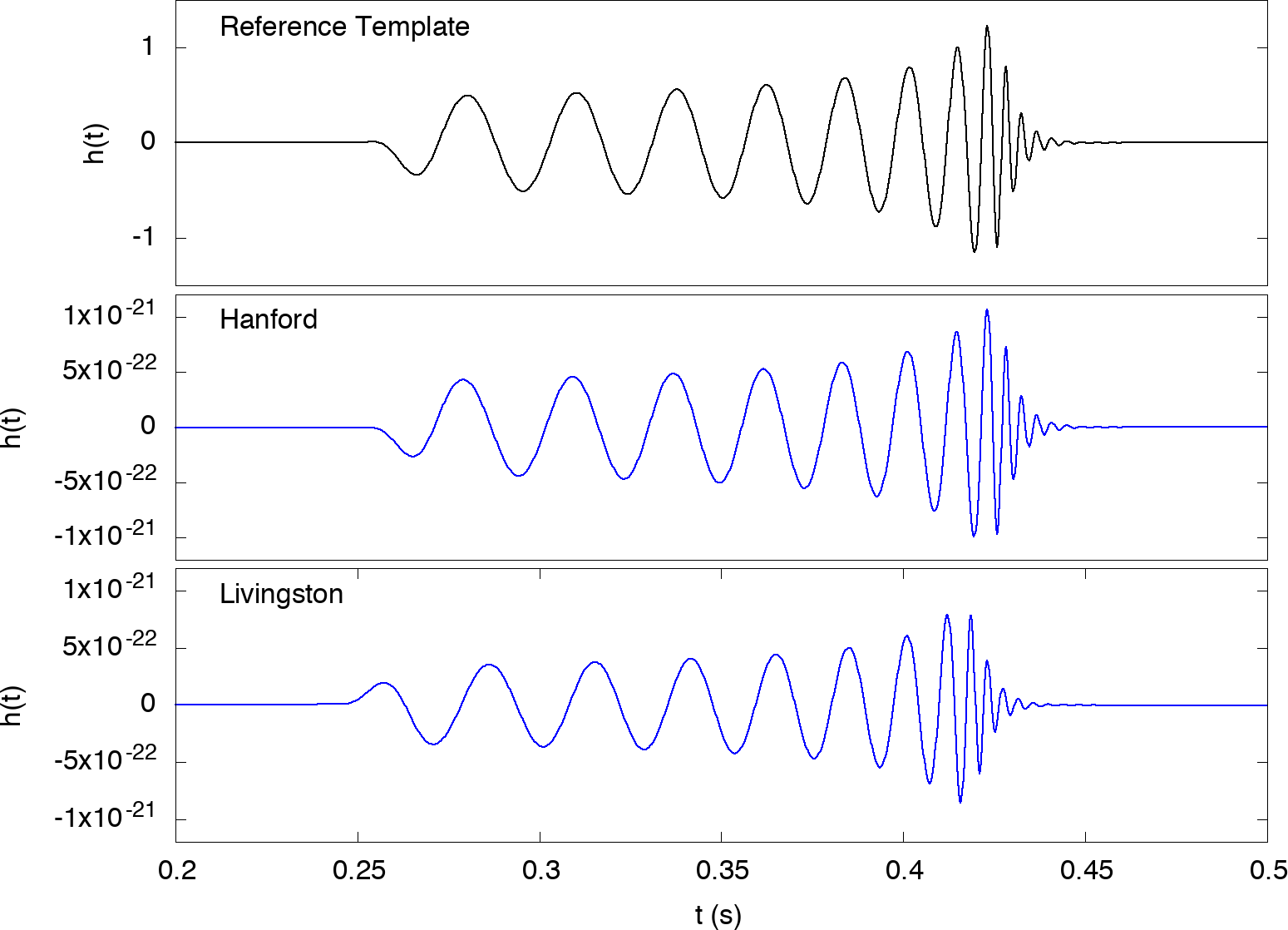}  
\caption{\label{fig:templates}  The reference numerical relativity template provided by the GWOSC~\cite{NRwaveform} for GW150914 is shown in the upper panel. The lower panels show the time, amplitude and phase shifted versions of the template that maximize the likelihood in each LIGO detector individually.}
\end{figure}

As introduced above, the physical parameters of the signal $\boldsymbol{\theta}$ determine the shape and amplitude of the gravitational-wave signal 
$h(t;\boldsymbol{\theta})$.
Numerical relativity simulations can be used to generate reference templates~\cite{LOSC} using intrinsic parameters taken from the Bayesian parameter estimation studies. However, the templates still need to be projected onto the detectors using an appropriate set of extrinsic parameters. In a single detector the projection is equivalent to time shifting, phase shifting and rescaling the reference template: $\tilde h(\alpha, \delta t, \delta \phi)(f) =  \alpha \, \tilde h_{\rm ref}(f) e^{2 \pi i f \delta t + i\delta \phi}$. 

Figure~\ref{fig:templates} shows the reference numerical relativity template from Figure 2 of the GW150914 discovery paper~\cite{Abbott:2016blz} along with maximum likelihood projections onto each detector.
A smooth taper has been applied to the start of the template to avoid spectral leakage when transforming to the Fourier domain. 
The data file for the template was taken from the original posting at the GWOSC~\cite{NRwaveform} and originates from the simulation SXS:BBH:0305, calculated for a system with a mass ratio of $q=0.819$,  spins aligned with the orbital angular momentum with dimensionless magnitudes $\chi_1=0.330$ and $\chi_2=-0.440$, and detector-frame total mass scaled to $M=74.6 M_\odot$.
These waveform parameters are consistent with those eventually determined for GW150914, within uncertainties, but do not exactly maximize the likelihood globally.
Using the maximization procedure described in Section~\ref{s:searches} one finds that the signal arrived at the LIGO-Livingston detector 7.08 ms before the LIGO-Hanford detector, had a larger amplitude projected onto the antenna response pattern in LIGO-Hanford by a factor of 1.24, and had a phase difference of $-2.9$ radians.  
These are, however, based on finding maximum-likelihood matches to the detector data individually with a fixed waveform without constraining them to be consistent (for example, the relative time shift could in principle be greater than the maximum light travel time between the detectors), a simplified procedure compared to the simultaneous multi-detector likelihood maximization described in Section~\ref{s:pe}.
When a loud signal is present in the data the individual and joint maximization techniques yield consistent results.

In Figure 1 of the GW150914 discovery paper~\cite{Abbott:2016blz} the LIGO-Hanford data were inverted (corresponding to a phase shift of $\pm \pi$) and overlaid on the LIGO-Livingston data to illustrate the similarity of the signals in the two detectors with minimal processing of the raw data.
%
%
In addition, the reference numerical relativity template described above was approximately matched to the LIGO-Hanford and LIGO-Livingston data by adjusting the relative phase, amplitude and time offset.  These adjusted templates for each detector were passed through the same bandpass and notch (band-reject) filters as the data and were then subtracted to produce the residuals plotted in the third row of Figure 1 in that paper.  Because those ``Fig 1 PRL'' residuals were not globally optimized and were calculated from filtered data, they produce a somewhat different result than minimizing the residuals in the whitened and bandpassed data, as we will see below.

\begin{figure}[h]
\centering \includegraphics[scale=0.3]{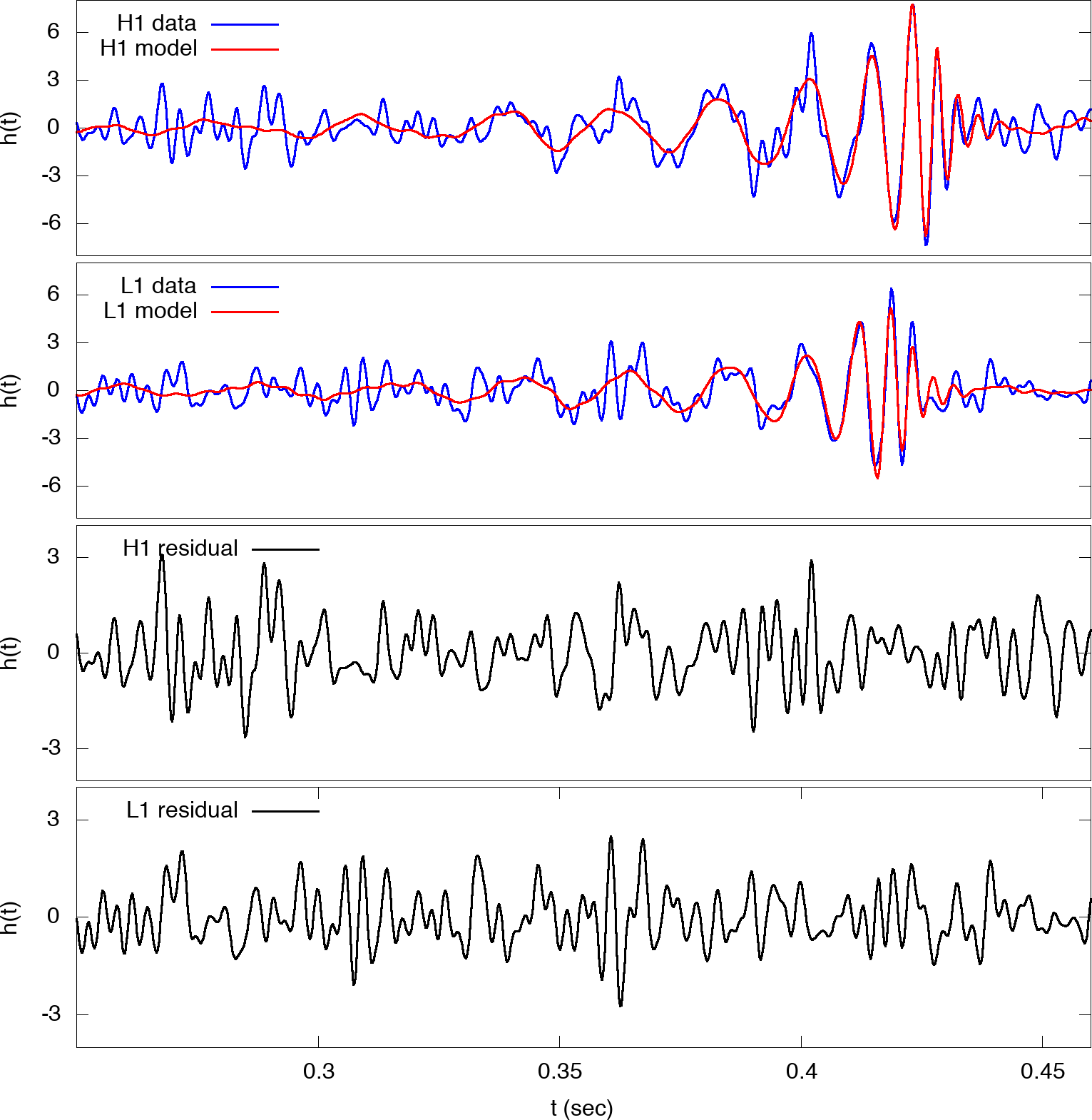}  
\caption{\label{fig:HL} The upper panels show the whitened and bandpassed data in the LIGO-Hanford and LIGO-Livingston detectors relative to GPS time 1126259462. The maximum likelihood whitened templates have been superimposed on the data. The lower panels show the residuals that are produced by subtracting the templates from the data.}
\end{figure}

Figure~\ref{fig:HL} compares the whitened data to the whitened numerical relativity templates, maximized over arrival time, amplitude and phase, in the LIGO-Hanford and LIGO-Livingston detectors. Also shown are the residuals produced by subtracting the templates from the data. Prior to the publication of the GW150914 discovery paper~\cite{Abbott:2016blz}, multiple tests were applied to the residuals to verify they were consistent with noise. The whitened residuals in each detector were found to be consistent with a Gaussian distribution: the Fourier amplitudes pass the Anderson-Darling test (see Figure~\ref{fig:pe} in Section~\ref{s:pe}), and the Fourier phases were found to be randomly distributed. The residuals from Bayesian parameter estimation studies~\cite{TheLIGOScientific:2016wfe} were analyzed using a wavelet reconstruction algorithm~\cite{Cornish:2014kda} that is able to detect coherent signals of general morphology. The degree of coherence in the GW150914 residuals was found to be entirely consistent with noise~\cite{TheLIGOScientific:2016src}. 

\subsection{Correlation analyses} \label{ss:corranal}

A simpler, though less sensitive, test for coherence is to cross-correlate the data in the two detectors. 
The cross-correlation can be computed either in the time domain or the frequency domain using the whitened residuals. The correlation in the time domain is defined as:
\begin{equation}
C(\tau) =\frac{ \int H(t-\tau) L(t) dt}{\sqrt{\int H^2(t)  dt \int L^2(t)  dt }} ~,
\end{equation}
where $H(t)$ and $L(t)$ represent the data streams from LIGO-Hanford and LIGO-Livingston respectively.
When working with a finite data segment of duration $T$ the data may be taken to be periodic: $H(t) = H(t+T)$. The correlation measure is very sensitive to the positioning and duration of the time window and the bandpass filtering that is applied to the data. To make meaningful statements about the significance of the correlation we need to know the distribution of the correlation measure for uncorrelated white noise, and these distributions change depending on the duration and bandpass.
When applied to uncorrelated, unit variance Gaussian noise, the correlation coefficients follow a zero mean Gaussian distribution with a variance that depends on the duration and bandpass.
Following~\cite{Creswell:2017rbh} we apply the correlation analysis to four different time windows. 
The standard deviations for white Gaussian noise are $\sigma=0.0870$ for the 0.2 second segment, $\sigma = 0.121$ for either 0.1 second segment and $\sigma=0.193$ for the 40 ms segment.

\begin{figure}[h]
\centering \includegraphics[scale=0.8]{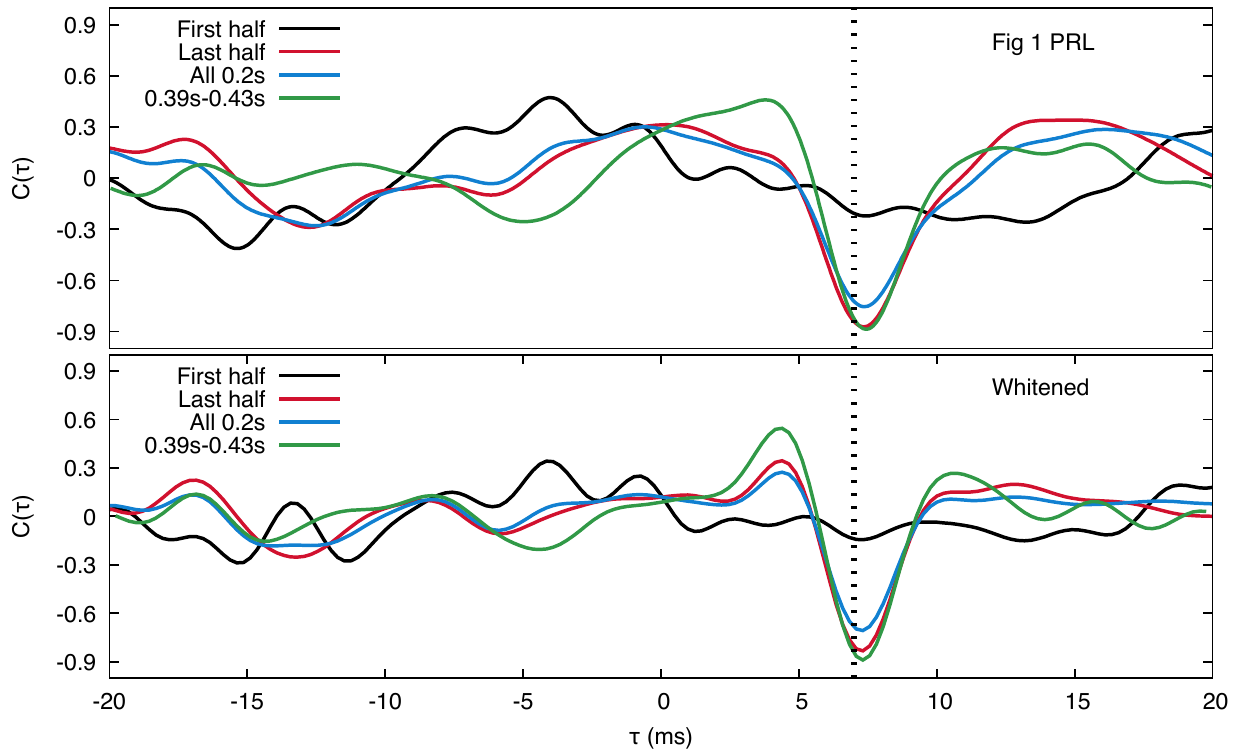} 
\caption{\label{fig:corrdata} Correlations between the LIGO-Hanford and LIGO-Livingston detector data using the same four time intervals used in the analysis in~\cite{Creswell:2017rbh}. One time interval covers the full 0.2 seconds of data shown in Figure~\ref{fig:HL}, and two others cover the first and last 0.1 seconds of the data. In addition, a very short time interval of duration 40 ms was selected that covers the peak of the signal. A time lag of 7 ms is highlighted as a dotted vertical line. The upper panel uses the filtered data from Figure 1 of the GW150914 discovery paper~\cite{Abbott:2016blz}, while the lower panel uses the whitened data shown in Figure~\ref{fig:HL}. }
\end{figure}

Figure~\ref{fig:corrdata} shows the correlations  using the whitened data shown in Figure~\ref{fig:HL} (bottom panel), and in addition, the bandpass/notch-filtered data used to produce the panels 
in Figure 1 of the GW150914 discovery paper~\cite{Abbott:2016blz} (top panel). There is a clear anti-correlation peak in the LIGO-Hanford -- LIGO-Livingston data at a time lag of $\sim 7.3$ ms, which is consistent with the time delay inferred for the gravitational-wave signal.

\begin{figure}[h]
\centering  \includegraphics[scale=0.8]{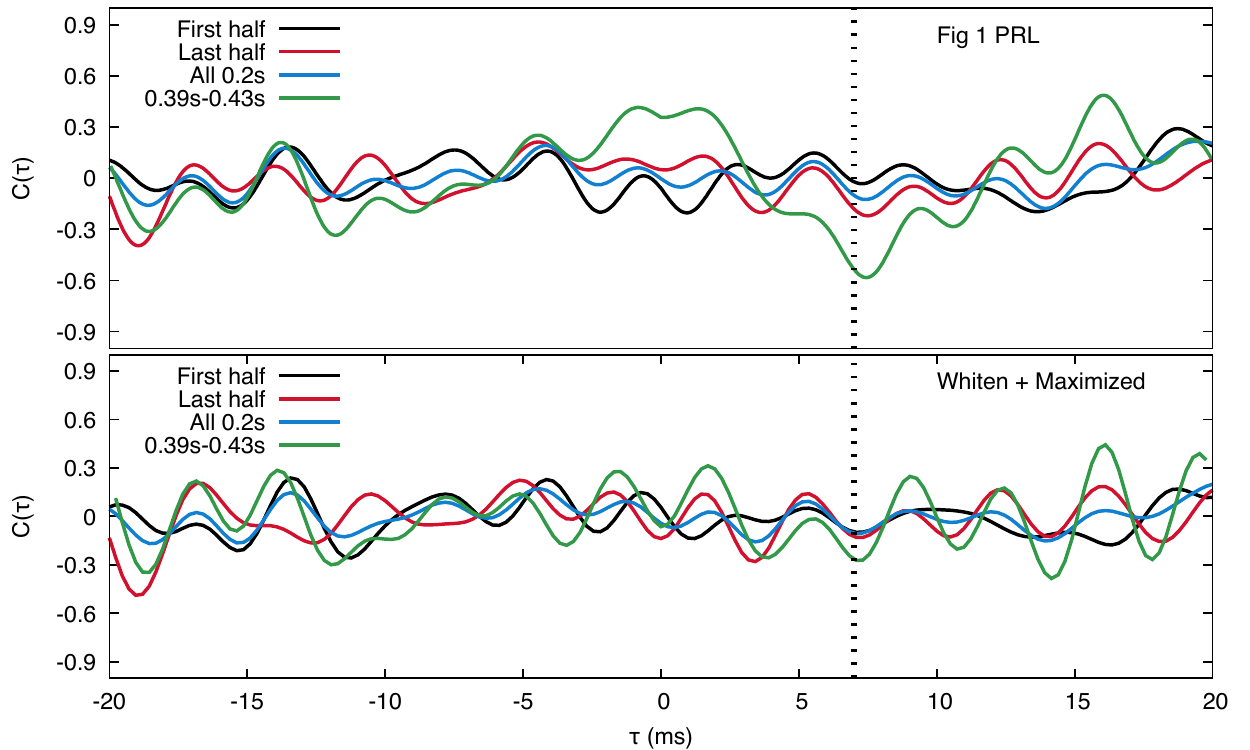}  
\caption{\label{fig:corrres} Correlations between the LIGO-Hanford and LIGO-Livingston residuals using the same four time intervals as Figure~\ref{fig:corrdata}.
The upper panel uses the residuals shown in Figure 1 of the GW150914 discovery paper~\cite{Abbott:2016blz}, while the lower panel uses the whitened residual time series shown in Figure~\ref{fig:HL}. The whitened residuals from the maximum likelihood signal subtraction show no significant correlations at any time lag for any of the time windows.}
\end{figure}

In contrast, Figure~\ref{fig:corrres} shows the correlations in the residuals produced using the procedure described above. The residuals from Figure~\ref{fig:HL} show no notable anti-correlation at $\sim 7$ ms (bottom panel), while those from Figure 1 of the GW150914 discovery paper~\cite{Abbott:2016blz} have a slight dip at this time lag (top panel), reflecting the fact that the reference waveform used for illustration in that paper was not the maximum likelihood waveform. For the shortest integration interval, the residuals from Figure 1 of the GW150914 discovery paper have a $\sim 3$ sigma anti-correlation at a time lag of 
$\sim 7.45$ ms, which while marginally consistent with noise, is evidence that the signal subtraction was imperfect. In contrast, the residuals produced using the amplitude/time/phase maximized NR waveforms and whitened data show no significant excursions, and are fully consistent with noise. This is also the case for the residuals from the Bayesian parameter estimation described in Section~\ref{s:pe}. Independent analyses of the GW150914 data have also found no significant correlations between the residuals in the Hanford and Livingston detectors~\cite{Green:2017voq,Nielsen:2018bhc}.

%% file: conclusion_new.tex
\section{Conclusions}
\label{sec:conclusions}
In this article we presented a description of the properties of data from the LIGO and Virgo detectors, and an overview of the analysis methods used by the LVC in identifying and characterizing gravitational-wave signals from the coalescence of binary black hole and binary neutron star systems.
We have especially looked closely at the data surrounding the first detection, GW150914~\cite{Abbott:2016blz,LOSC:GW150914,LIGOScientific:2018mvr}. 
Contrary to the claims made in~\cite{Creswell:2017rbh}, there are no anomalous or unexpected correlations to be seen in association with the observed gravitational-wave events~\cite{LIGOScientific:2019fpa}, including GW150914~\cite{LOSC:GW150914}. Other analyses by independent researchers have come to similar conclusions about the correctness of the LIGO-Virgo results~\cite{Green:2017voq,Nielsen:2018bhc,Roulet:2018jbe,Nitz:2018imz}. 

Proper handling of the LIGO and Virgo data is critical for conducting an analysis correctly. 
As an example, in this paper we have used the whitened maximum likelihood waveforms (as described in Sections~\ref{s:pe} and ~\ref{sec:res}) for GW150914, which when subtracted from the data, produce residuals that are consistent with Gaussian noise, 
and show no correlation between different detectors.
If the template waveforms subtracted from the data are not sufficiently good matches to the real gravitational-wave signal, then a remainder of that signal will survive in the resulting residuals, which may thus exhibit nontrivial correlations.


Figure 1 of~\cite{Abbott:2016blz} was constructed to show as simply as possible that the signal is compatible with general relativity.  It does not illustrate the full LSC-Virgo statistical data analysis.
The figure was described in~\cite{Abbott:2016blz} as a visualization of the gravitational-wave signal at the LIGO detectors and a comparison to one numerical relativity waveform which is consistent with the gravitational-wave data. A statistical claim about the numerical relativity waveform
and the residuals of Figure 1 of \cite{Abbott:2016blz} was not intended, although unfortunately the figure may have been interpreted in that way.

The LVC conducted extensive statistical studies of the GW150914 signal and of the surrounding noise, which are documented in~\cite{TheLIGOScientific:2016wfe}. Note that a whitened time series of GW150914 was presented in the parameter estimation companion paper for the discovery; see Figure 6 of~\cite{TheLIGOScientific:2016wfe}. Those studies, as well as the simpler investigations given here, support the interpretation that the signal is well matched by a black hole merger solution of general relativity. 
The validity of this conclusion has been supported by subsequent data and analysis by the LVC (including studies on all binary black hole produced gravitational-wave signals detected in observing runs O1 and O2~\cite{LIGOScientific:2019fpa}) as well as independent analyses. 

The gravitational-wave data for Advanced LIGO and Advanced Virgo can be characterized as locally stationary and Gaussian, with deviations when glitches are present. 
The LVC conducts extensive data quality, detector characterization, and calibration studies in order to be confident of the reported detections~\cite{TheLIGOScientific:2016zmo,TheLIGOScientific:2017lwt,Abbott:2016jsd,Acernese:2018bfl}.

However, it is not necessary to assume that the data are stationary and Gaussian to search for, and to detect with high confidence, gravitational waves from compact binary coalescence.
Instead, LIGO-Virgo searches for gravitational waves use various methods to estimate the false alarm rate directly from the data, for example, by introducing a relative time shift between the detectors.

Previous studies have also demonstrated that the LVC's parameter estimation results are reliable~\cite{Raymond:2009cv,vanderSluys:2009bf,vanderSluys:2007st,Berry:2014jja, Abbott:2016wiq, Pankow:2018qpo}. The parameter estimation routines were also robust for the gravitational waves from the binary neutron star merger GW170817 where there was a noise glitch in the LIGO-Livingston data overlapping with the gravitational-wave signal~\cite{TheLIGOScientific:2017qsa,Pankow:2018qpo}. 
Parameter estimates obtained by researchers outside the LVC for GW170817 are comparable with, and support the conclusions of, the LVC analyses~\cite{Dai:2018dca,Radice:2018ozg,De:2018uhw}; these studies were made possible by the public release of the gravitational-wave data~\cite{LOSC}.

While the examples in this paper have concentrated on the events GW150914 and GW170817, the conclusions presented have been demonstrated to be valid for the analysis of the data containing all 11 gravitational-wave events detected by LIGO and Virgo to date~\cite{LIGOScientific:2018mvr,LIGOScientific:2019fpa}.
As the LIGO and Virgo collaborations report more events~\cite{LIGOScientific:2018mvr,Aasi:2013wya}, independent analyses of the data associated with these events by the broader scientific community will be highly valuable and may well produce new insights. To this end, in this paper we have tried to provide some guidance on the nature of LIGO and Virgo detector noise and on the extraction of gravitational-wave signals. The LVC encourages the scientific community to analyze its data; LIGO and Virgo data will continue to be made publicly available on the GWOSC website~\cite{LOSC}.

%% file: LVCacknowledgments.tex
The authors gratefully acknowledge the support of the United States
National Science Foundation (NSF) for the construction and operation of the
LIGO Laboratory and Advanced LIGO as well as the Science and Technology Facilities Council (STFC) of the
United Kingdom, the Max-Planck-Society (MPS), and the State of
Niedersachsen/Germany for support of the construction of Advanced LIGO 
and construction and operation of the GEO600 detector. 
Additional support for Advanced LIGO was provided by the Australian Research Council.
The authors gratefully acknowledge the Italian Istituto Nazionale di Fisica Nucleare (INFN),  
the French Centre National de la Recherche Scientifique (CNRS) and
the Foundation for Fundamental Research on Matter supported by the Netherlands Organisation for Scientific Research, 
for the construction and operation of the Virgo detector
and the creation and support  of the EGO consortium. 
The authors also gratefully acknowledge research support from these agencies as well as by 
the Council of Scientific and Industrial Research of India, 
the Department of Science and Technology, India,
the Science \& Engineering Research Board (SERB), India,
the Ministry of Human Resource Development, India,
the Spanish  Agencia Estatal de Investigaci\'on,
the Vicepresid\`encia i Conselleria d'Innovaci\'o, Recerca i Turisme and the Conselleria d'Educaci\'o i Universitat del Govern de les Illes Balears,
the Conselleria d'Educaci\'o, Investigaci\'o, Cultura i Esport de la Generalitat Valenciana,
the National Science Centre of Poland,
the Swiss National Science Foundation (SNSF),
the Russian Foundation for Basic Research, 
the Russian Science Foundation,
the European Commission,
the European Regional Development Funds (ERDF),
the Royal Society, 
the Scottish Funding Council, 
the Scottish Universities Physics Alliance, 
the Hungarian Scientific Research Fund (OTKA),
the Lyon Institute of Origins (LIO),
the Paris \^{I}le-de-France Region, 
the National Research, Development and Innovation Office Hungary (NKFIH), 
the National Research Foundation of Korea,
Industry Canada and the Province of Ontario through the Ministry of Economic Development and Innovation, 
the Natural Science and Engineering Research Council Canada,
the Canadian Institute for Advanced Research,
the Brazilian Ministry of Science, Technology, Innovations, and Communications,
the International Center for Theoretical Physics South American Institute for Fundamental Research (ICTP-SAIFR), 
the Research Grants Council of Hong Kong,
the National Natural Science Foundation of China (NSFC),
the Leverhulme Trust, 
the Research Corporation, 
the Ministry of Science and Technology (MOST), Taiwan
and
the Kavli Foundation.
The authors gratefully acknowledge the support of the NSF, STFC, INFN and CNRS for provision of computational resources.
This article has been assigned the document number LIGO-P1900004.